\documentclass{nhmarxiv}
  \usepackage{amsmath}
  \usepackage{amssymb}
  \usepackage{paralist}
  \usepackage{graphics} 
  \usepackage{epsfig} 

  \usepackage{hyperref} 

\def\currentvolume{X}
 \def\currentissue{X}
  \def\currentyear{200X}
   \def\currentmonth{XX}
    \def\ppages{X--XX}

\theoremstyle{definition}

\title[Synchronization in the World Trade Web]
      {Modeling international crisis synchronization in the World Trade Web}

\author[Pau Erola, Albert D\'{\i}az-Guilera, Sergio G\'omez and Alex Arenas]{}


\email{pau.erola@urv.cat}

\begin{document}
\maketitle

\centerline{\scshape Pau Erola }
\medskip
{\footnotesize
 \centerline{Departament d'Enginyeria Inform\`atica i Matem\`atiques, Universitat Rovira i Virgili}
   \centerline{Pa\"{\i}sos Catalans 26, 43007 Tarragona (Spain)}
} 

\medskip

\centerline{\scshape Albert D\'{\i}az-Guilera }
\medskip
{\footnotesize
 \centerline{Departament de F\'{\i}sica Fonamental, Universitat de Barcelona}
   \centerline{Diagonal 647, 08028 Barcelona (Spain)}
} %

\medskip

\centerline{\scshape Sergio G\'omez and Alex Arenas }
\medskip
{\footnotesize
 \centerline{Departament d'Enginyeria Inform\`atica i Matem\`atiques, Universitat Rovira i Virgili}
   \centerline{Pa\"{\i}sos Catalans 26, 43007 Tarragona (Spain)}
} %

\bigskip

 \centerline{(Communicated by the associate editor name)}


\begin{abstract}
Trade is a fundamental pillar of economy and a form of social organization. Its empirical characterization at the worldwide scale is represented by the World Trade Web (WTW), the network built upon the trade relationships between the different countries. Several scientific studies have focused on the structural characterization of this network, as well as its dynamical properties, since we have registry of the structure of the network at different times in history. In this paper we study an abstract scenario for the development of global crises on top of the structure of connections of the WTW. Assuming a cyclic dynamics of national economies and the interaction of different countries according to the import-export balances, we are able to investigate, using a simple model of pulse-coupled oscillators, the synchronization phenomenon of crises at the worldwide scale. We focus on the level of synchronization measured by an order parameter at two different scales, one for the global system and another one for the mesoscales defined through the topology. We use the WTW network structure to simulate a network of Integrate-and-Fire oscillators for six different snapshots between years 1950 and 2000. The results reinforce the idea that globalization accelerates the global synchronization process, and the analysis at a mesoscopic level shows that this synchronization is different before and after globalization periods: after globalization, the effect of communities is almost inexistent.
\end{abstract}


\section{Introduction}

In September 2008, the bankruptcy of Lehman Brothers marked for many the beginning of the current global crisis. In an increasingly globalized world, the financial crisis spread relentlessly. Recent theories of financial fragility link globalization with economic cycles, i.e.\ when local crises coincide with bad credit regulation and failures in international monetary arrangements. The globalization process in recent years has been accelerated due to to the increase of international trade. The study of financial crises has always attracted a fair amount of interest, but we still know very little about them. Minsky \cite{minsky1992,minsky1986} defined financial crises as a natural consequence of changes in the economic cycle and the fragility of the structure of debt. In a capitalist economy the financial system swings between robustness and fragility and does not rely upon exogenous shocks to generate business cycles of varying severity. The crisis is triggered when concur falls in economic activity, poor management of bank credit and systemic liquidity requirements.

According to this definition, a global crisis would be highly unusual, and will only occur when economic cycles of all economies are synchronized. However, in September 2008 the bankruptcy of Lehman Brothers spread the North American financial crisis relentlessly world wide. The magnitude of this financial crisis is hard to explain and we will sustain our work on theories that point to a possible relationship between the globalization process and the synchronization of economic cycles.

Deardorff \cite{deardorff} defines globalization as ``\emph{the increasing world-wide integration of markets for goods, services and capital}'', stressing the idea that globalization is affecting the world as a whole. Global trade and financial flows have been increasing along the last century; but around mid eighties it has been observed an acceleration of this process which has been identified with the appearance of global trade. Free trade and increased financial flows can set the channels to spill shocks world wide. Kose \emph{et al.} \cite{kose2008} have analyzed the evolution of business cycles over the globalization period 1985-2005, finding some convergence of business cycle fluctuations among industrial economies and among emerging market economies. Instead, they find little changes in the degree of international synchronization.

When focusing on real data, there is also evidence of synchronization phenomena of economic cycles in the WTW \cite{li03}. Using correlation trade-data analysis, and the substrate of the WTW the authors show the presence of strong correlated synchronized behavior, without imposing any dynamics on the system. The paradigmatic synchronization of oscillators in complex networks, has attracted lot of attention in the physics literature \cite{physrep}. Nevertheless, there is no model of economic cycles in the WTW reporting such synchronization phenomenon.

Here we analyze how the effects of globalization can affect the ability of synchronization of business cycles. To this end, we use the network of international trade (WTW) \cite{wtwnet} where we will represent each country economical cycle with an Integrate-and-Fire oscillator \cite{mirollo1990}. We have chosen snapshots of this network at six different years between 1950 and 2000, some for the pre-globalization period (1950, 1960, 1970 and 1980) and the rest belonging to the globalization period (1990 and 2000). In an increasingly globalized world, we can expect that more tightly coupled countries synchronize quickly. We also analyze how this synchronization is driven, at the mesoscopic scale, by the existence of modules in the network representing stronger associations of trading between certain groups of nations.


\section{A simple model of economic cycles in the WTW}

We first present the WTW network and we overview some basic notation. WTW represents the international transactions among countries as a network. The network is formed by $N$ nodes, one for each country, and $L$ links that correspond to trade flows. Each link has an associated weight $w_{ij}$ that represents the trade between two countries $i$ and $j$. Thus, the link $w_{ij}$ accounts for the exports of node $i$ to node $j$, and $w_{ji}$ for the imports of $i$ from $j$. For a characterization of this network and his evolution see \cite{marian2003, galar07,fagiolo09,he10,squar11}.

For each node $i$ we can calculate the in-strength
\begin{equation}
  s^{\mbox{\scriptsize{in}}}_i = \sum_j w_{ji}\,,
\end{equation}
equivalent to the total imports of a country, and the out-strength
\begin{equation}
  s^{\mbox{\scriptsize{out}}}_i = \sum_j w_{ij}\,,
\end{equation}
equivalent to its total exports.

To model the economic cycle of each country we associate an oscillator to every node of the network. The interaction of these oscillators is performed through the commercial channels represented by the links of the network. In this article we use the simple Mirollo and Strogatz {\em Integrate-and-Fire Oscillators} (IFO) model \cite{mirollo1990}. In this model each oscillator is characterized by a monotonic increasing state variable $x\in[0,1]$ that evolves according to
\begin{equation}
  x=f(\phi)=\frac{1}{b} \ln(1+(e^b - 1)\phi)\,,
  \label{x}
\end{equation}
where $\phi\in[0,1]$ is a phase variable proportional to time, and $b$ is the dissipation parameter that measures the extent to which $x$ is concave down; when $b$ approaches zero $f$ becomes a linear function. We can calculate $\phi$ using the inverse function
\begin{equation}
  \phi=f^{-1}(x)=\frac{e^{bx}-1}{e^b-1}\,.
  \label{phi}
\end{equation}

When the variable $x$ attains the threshold $x=1$ it is said to {\em fire}, and it is instantly reset to zero, after which the cycle repeats. Now let us assume that a node $i$ of our network of oscillators fires. This node, in turn, transmits to all its neighbors $j$ an excitation signal of magnitude $\varepsilon_{ij}>0$, thus leaving their state variables $x^{+}_{j}$ with values
\begin{equation}
  x^{+}_{j} =
  \begin{cases}
    x_j+\varepsilon_{ij} & \text{if $x_j+\varepsilon_{ij} < 1$,}
    \\
    0 &\text{if $x_j+\varepsilon_{ij} \geqslant 1$.}
  \end{cases}
  \label{xplus}
\end{equation}
If $x_j+\varepsilon_{ij} \geqslant 1$, oscillator $j$ is also reset and propagates the fire signal to its neighbors, thus generating cascades of fires.

The abstraction we propose is that each country has its own economic cycle, which we assume evolves sooner or later towards a crisis; this is represented by an IFO. At this moment, the nation {\em fires}, i.e.\ propagates the problem to other countries through its connection in the WTW by boosting their own evolution to a crisis. WTW presents a large diversity in the economic weight of countries and their trade flow. It is reasonable to think that a shock in a small country is not spread to a large country with the same intensity than vice versa. And regardless of country size, the volume of transactions appears to be a factor of the ability of trade channel to synchronize the cycles of two economies. To reflect this dependence we set the excitation signal of node $i$ to its neighbors $j$ as
\begin{equation}
  \varepsilon_{ij}=\frac {w_{ji}}{s^{\mbox{\scriptsize{out}}}_j}\,,
  \label{epsij}
\end{equation}
which is the fraction of total exports of country $j$ going to the firing node $i$. For the sake of simplicity, we will assume identical oscillators, isolating the effect of synchronization of crises to the dynamics of interactions weighted by the trading exchanges.


\section{Simulation analysis}

The structure of the WTW, a densely connected weighted directed graph, makes certainly difficult to assess {\em a priori} the global outcome of different economic cycles interacting through pulses. No general theory exists to ascertain the stationary state, if any, of such a network of pulse-coupled oscillators. For this reason we will rely on computer simulations to shed some light on the behavior of this complex system. Previous studies of complex networks of pulse-coupled oscillators \cite{timme02} have shown the coexistence of synchronized and non synchronized dynamical regimes when the coupling $\varepsilon$ is positive (excitatory) in homogeneous and heterogeneous in degree networks. Also recurrent events of synchrony and asynchrony have been reported in complex networks when interactions have delay and refractoriness \cite{roth11}. Here we expect all this phenomenology to show up in our particular network, with essential differences between the snapshots at different years of the WTW.

We analyze six snapshots of the WTW corresponding to years between 1950 and 2000, with different values of $N$, the number of countries reported on the WTW. For the simulations, we assign an oscillator per country with random initial phase. The system evolves following the dynamics given by Eqs.~(\ref{x}) to~(\ref{xplus}), with $b=3$ and $\varepsilon$ as described in Eq.~(\ref{epsij}). The dynamics is stopped when 90\% of the oscillators are synchronized, and the statistics shown have been obtained after $10^3$ repetitions of this dynamics.

We will pay also attention to the modular structure of the different snapshots of the WTW detected using modularity \cite{newanaly,sizered}. Given a weighted directed network partitioned into communities, being $C_i$ the module to which node $i$ is assigned, the mathematical definition of modularity is expressed in terms of the weighted adjacency matrix $w_{ij}$, that represents the value of the weight in the link between nodes $i$ and $j$ ($0$ if no link exists), and the input and output strengths, $s^{\mbox{\scriptsize{in}}}_i$ and $s^{\mbox{\scriptsize{out}}}_i$ respectively, as
\begin{equation}
  Q=\frac{1}{2w}\sum_{i,j} \left(
        w_{ij}-\frac{s^{\mbox{\scriptsize{out}}}_i\, s^{\mbox{\scriptsize{in}}}_j}{2w}
      \right)\delta(C_i,C_j)\,,
\label{QW}
\end{equation}
where the Kronecker delta function $\delta(C_i,C_j)$ takes the values, 1 if node $i$ and $j$ are into the same module, 0 otherwise, and the total strength is $2w=\sum_{i,j} w_{ij}$. The modularity of a given partition is then, the probability of having edges falling within groups in the network minus the expected probability in an equivalent (null case) network with the same number of nodes, and edges placed at random preserving the nodes' strength. The larger the modularity the best the partitioning is, cause more deviates from the null case. Note that the optimization of the modularity cannot be performed by exhaustive search since the number of different partitions are equal to the Bell or exponential numbers \cite{bell}, which grow at least exponentially in the number of nodes $N$. Indeed, optimization of modularity is a NP-hard (Non-deterministic Polynomial-time hard) problem \cite{brandes}. As a consequence heuristics for the optimization of modularity \cite{newfast,clauset,rogernat,duch,pujol,newspect,meso} have become the only feasible (in computational time), reliable and accurate methods to detect modular structure up to now.

\begin{figure}[p]
  \begin{center}
    \begin{tabular}[t]{cc}
      \mbox{\includegraphics*[width=0.47\textwidth]{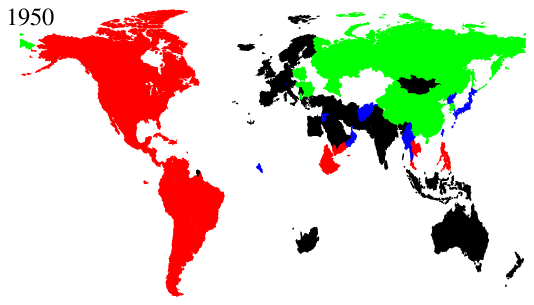}} &
      \mbox{\includegraphics*[width=0.47\textwidth]{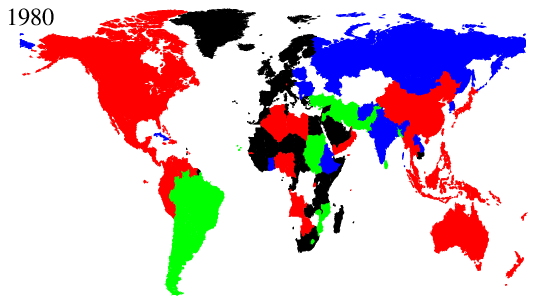}}
    \\
      \mbox{\includegraphics*[width=0.47\textwidth]{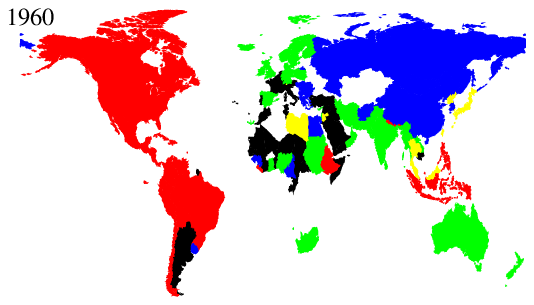}} &
      \mbox{\includegraphics*[width=0.47\textwidth]{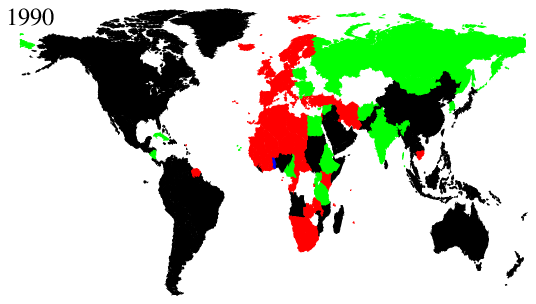}}
    \\
      \mbox{\includegraphics*[width=0.47\textwidth]{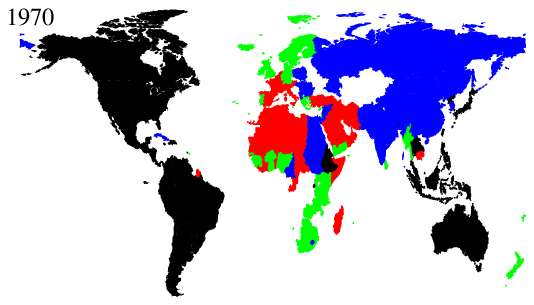}} &
      \mbox{\includegraphics*[width=0.47\textwidth]{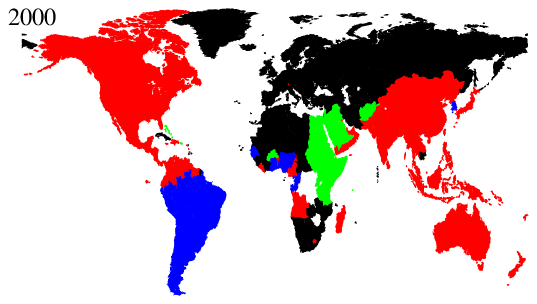}}
    \end{tabular}
  \end{center}
\caption{Topological communities found by modularity maximization. Each color corresponds to a different community. Since the community structure evolves through the years, the color of a certain country may change through time. Regions in white correspond to countries without information in the dataset, or to regions belonging to other countries.}
  \label{fig:comm1}
\end{figure}

The modularity analysis of the mesoscopic scale of the WTW reveals different topological communities for the different snapshots we analyzed. This is eventually an expected fact because the trading weights have changed over years, and consequently the topology of the WTW. In Fig.~\ref{fig:comm1} we show the world-map labeled according to the best partition into communities found, for each year, using modularity (see appendix for the detailed list of countries in each community). In Table~\ref{wtw} we present the number of communities, their size in number of countries, and their relative strength according to import-export data. The mesoscopic structure will be used to investigate the effects of communities in the proposed dynamics of crises spreading using the IFO model.

For the evaluation of the interaction dynamics, we follow the classical analysis of the order parameter $r$ that indicates the degree of synchronization of a system with $N$ oscillators. We recall that in our setup, synchronization will reflect the scope of the crisis at the international level. The order parameter is
\begin{equation}
  r=\left| \frac{1}{N} \sum_{j=1}^{N} e^{2\pi i \phi_{j}}\right|
      =\sqrt{ \left(\frac{1}{N}\sum_{j=1}^{N} \cos(2\pi \phi_{j})\right)^2 +
              \left(\frac{1}{N}\sum_{j=1}^{N} \sin(2\pi \phi_{j})\right)^2}\,,
  \label{r}
\end{equation}
which only depends on the phases $\phi_{j}$ of the oscillators. This parameter $r$ is 1 for complete phase synchronization, and close to 0 in the incoherent state.


\section{Results and discussion}

First, we examine the dynamics of the IFO model at the global (worldwide) scale, and after we fix our attention to the evolution of the same observables at the mesoscopic scale. The results support the idea that as the WTW evolves towards a more globalized trading, the topological mesoscopic structure is less and less representative of particularly different dynamical behavior, being the global phenomenology equivalent to the mesoscopic one, and vice versa.

\begin{figure}[p]
  \begin{center}
    \begin{tabular}[t]{cc}
      \mbox{\includegraphics*[width=0.47\textwidth]{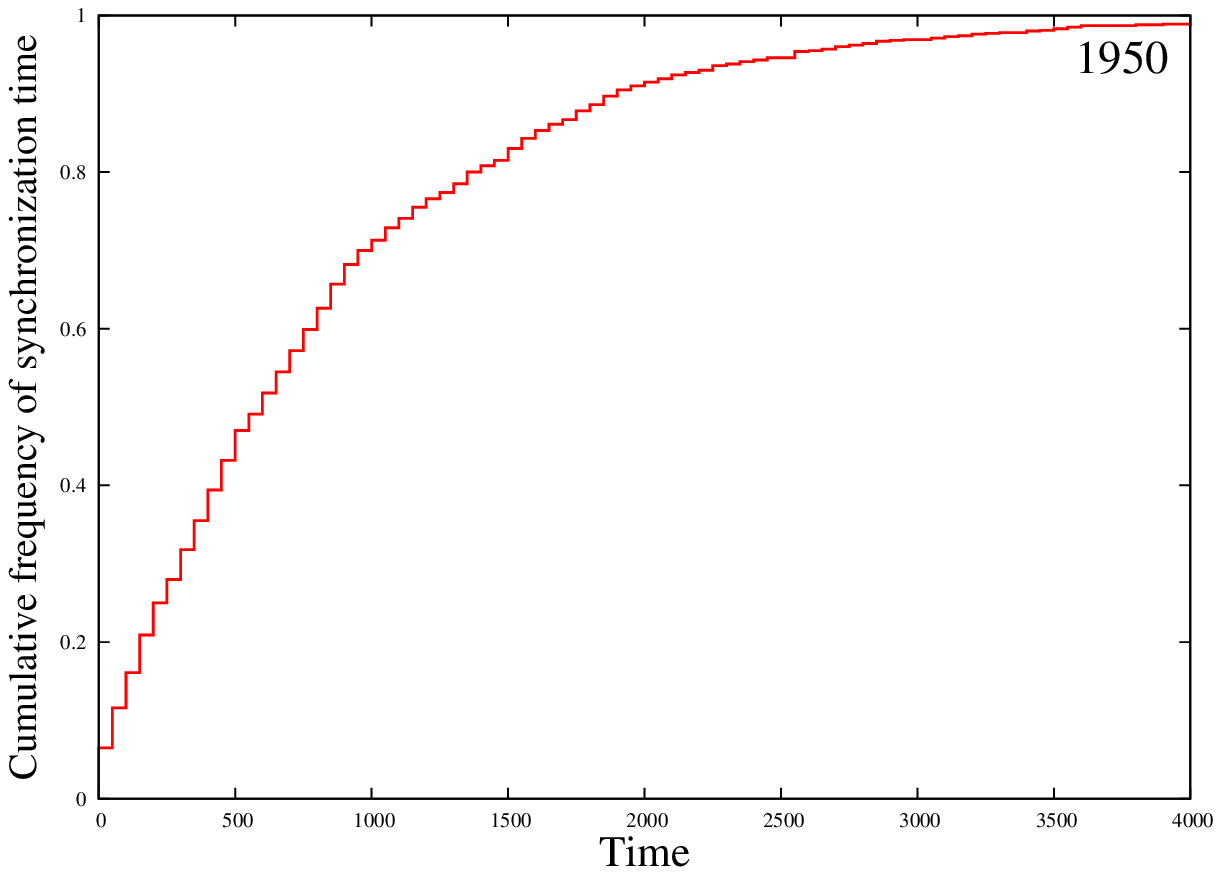}} &
      \mbox{\includegraphics*[width=0.47\textwidth]{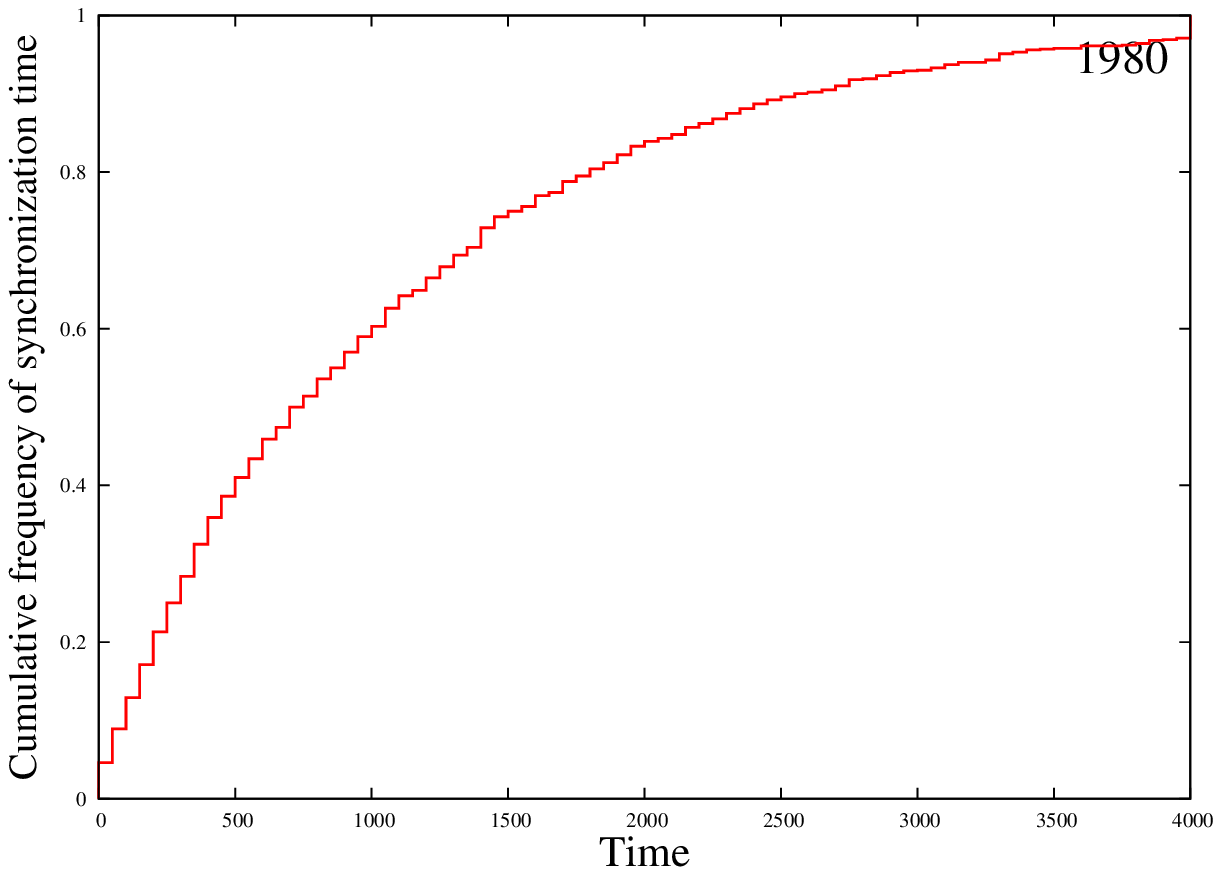}}
    \\
      \mbox{\includegraphics*[width=0.47\textwidth]{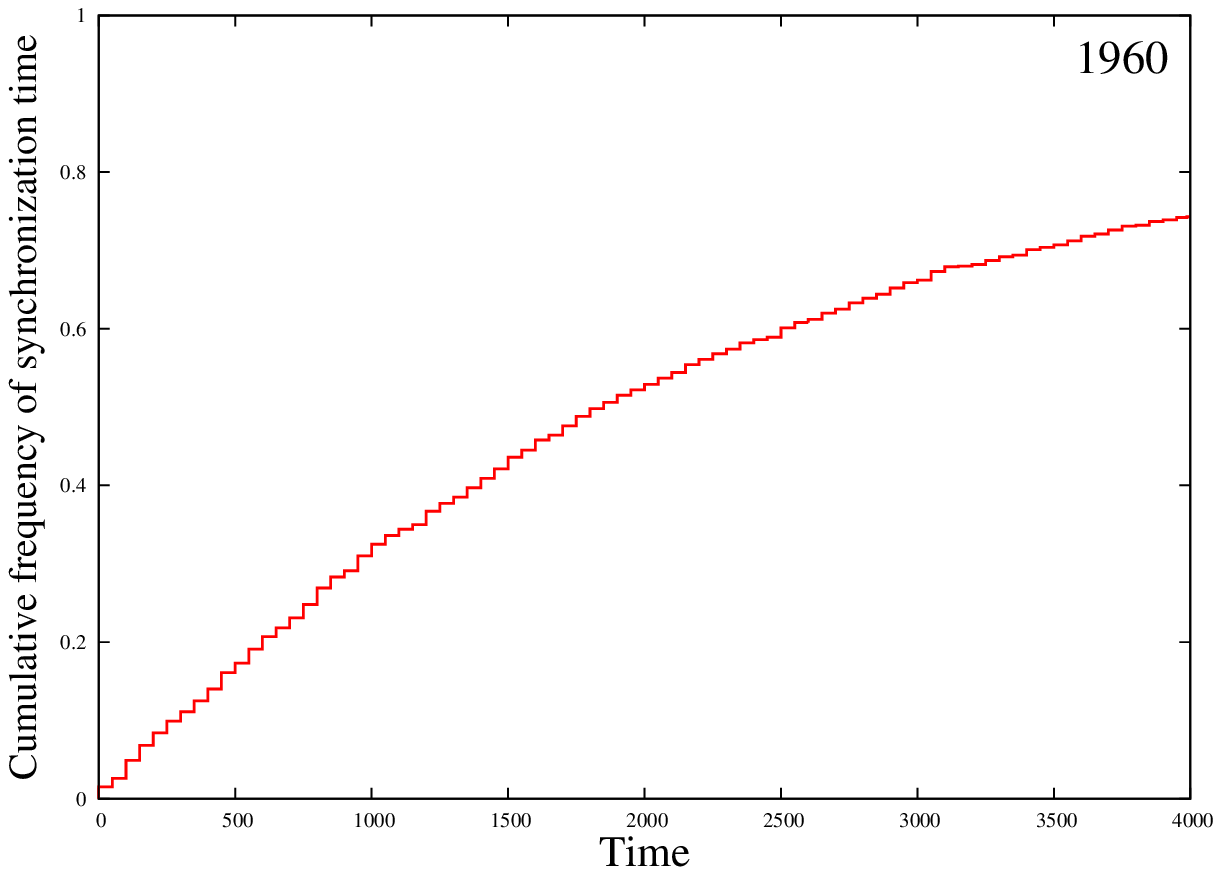}} &
      \mbox{\includegraphics*[width=0.47\textwidth]{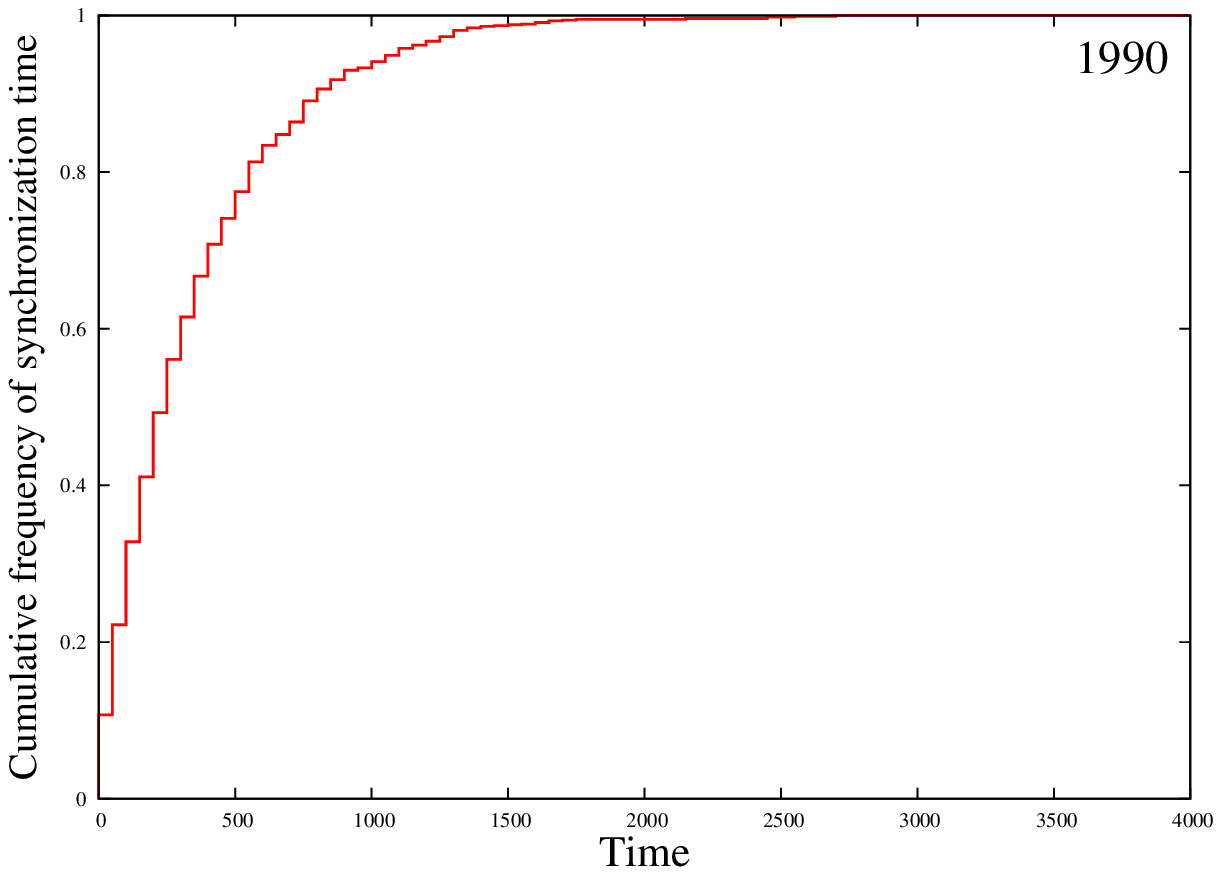}}
    \\
      \mbox{\includegraphics*[width=0.47\textwidth]{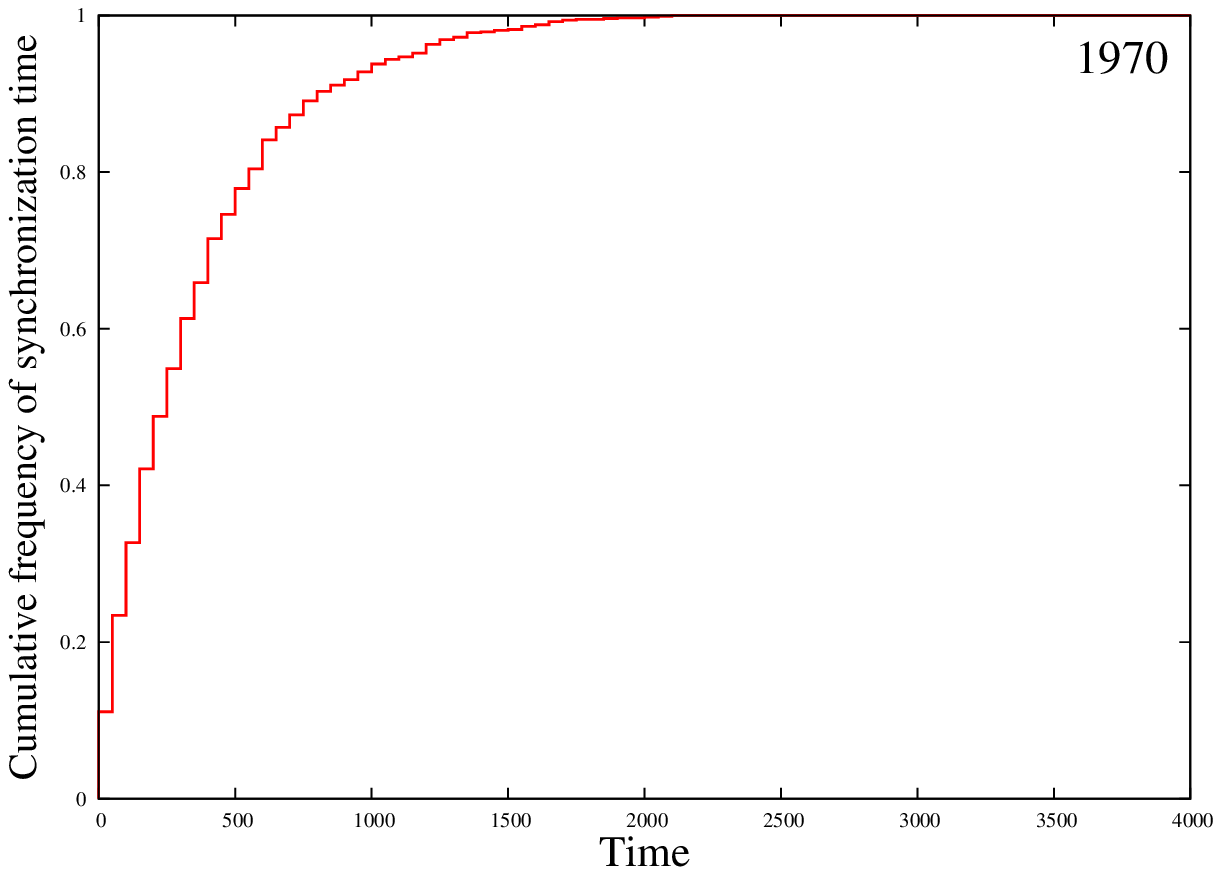}} &
      \mbox{\includegraphics*[width=0.47\textwidth]{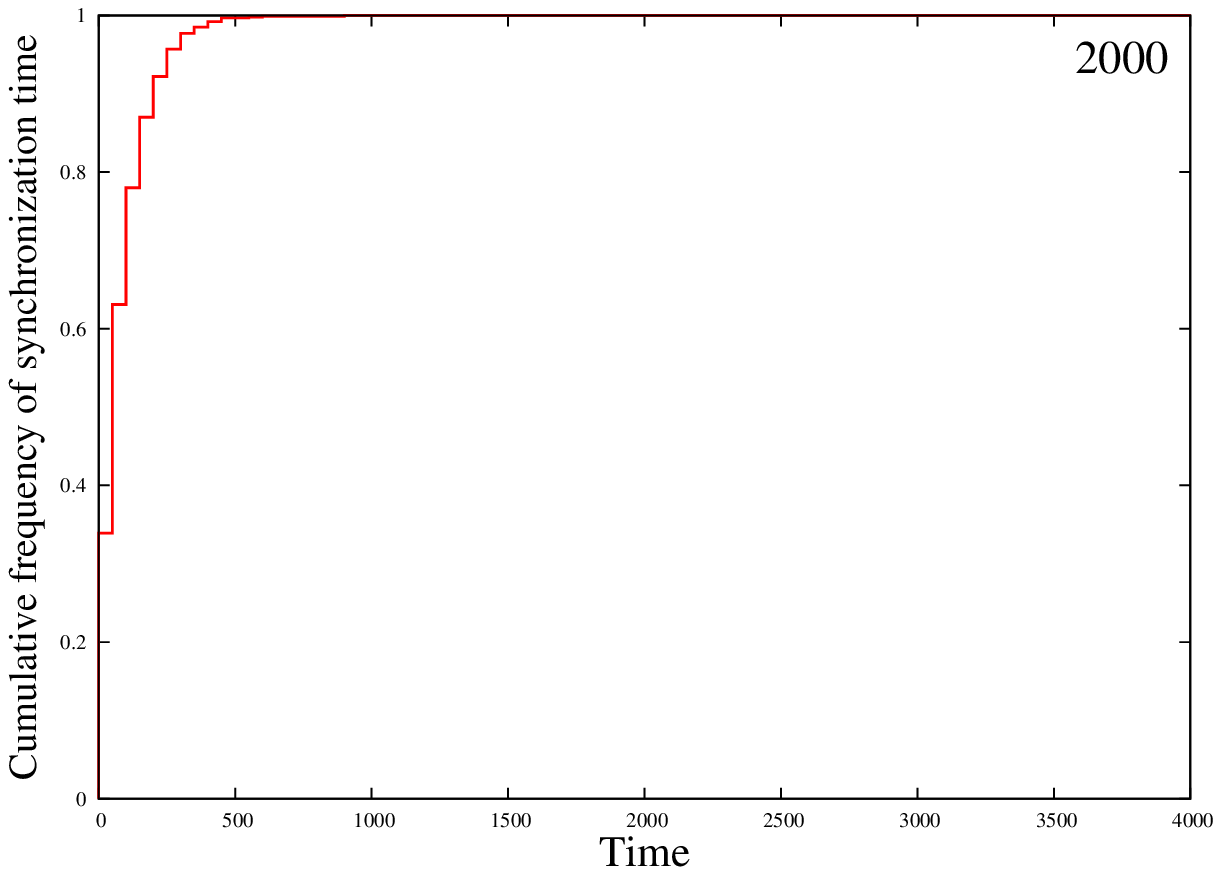}}
    \end{tabular}
  \end{center}
\caption{Cumulative frequency of synchronization time, i.e.\ the time needed to synchronize 90\%~of the nodes, for $10^3$~repetitions of the IFO dynamics. The width of the bins in these histograms is of 50~time cycles.}
  \label{fig:timeh}
\end{figure}

The first global aspect we investigate is that of the time needed for the whole network to synchronize and the scope of a certain firing event on the rest of the network, we will call it cascade or avalanche. To this end we first track the time needed to synchronize 90\% of nodes of the network. In Fig.~\ref{fig:timeh} we represent the cumulative distribution of time up to synchronization. Interestingly, the distribution shows that as the WTW evolves along years, the time for almost global synchronization is faster. Specially interesting is the change of slopes observed between 1960--1970 and 1980. From the data obtained, our analysis suggests that the globalization process should take place within this two decades, and producing a clear differentiation between the period before 1950--1960 and the posterior 1990--2000.

\begin{figure}[p]
  \begin{center}
    \begin{tabular}[t]{cc}
      \mbox{\includegraphics*[width=0.47\textwidth]{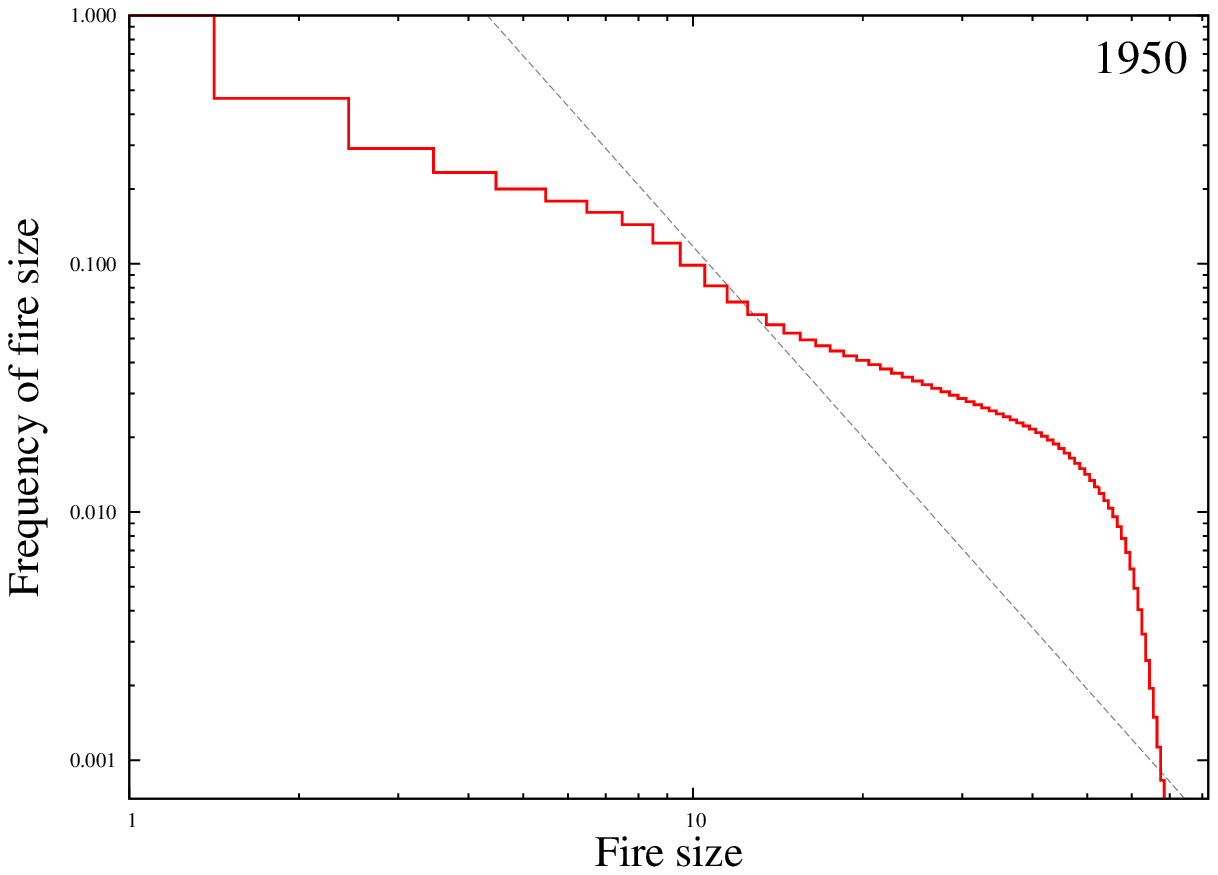}} &
      \mbox{\includegraphics*[width=0.47\textwidth]{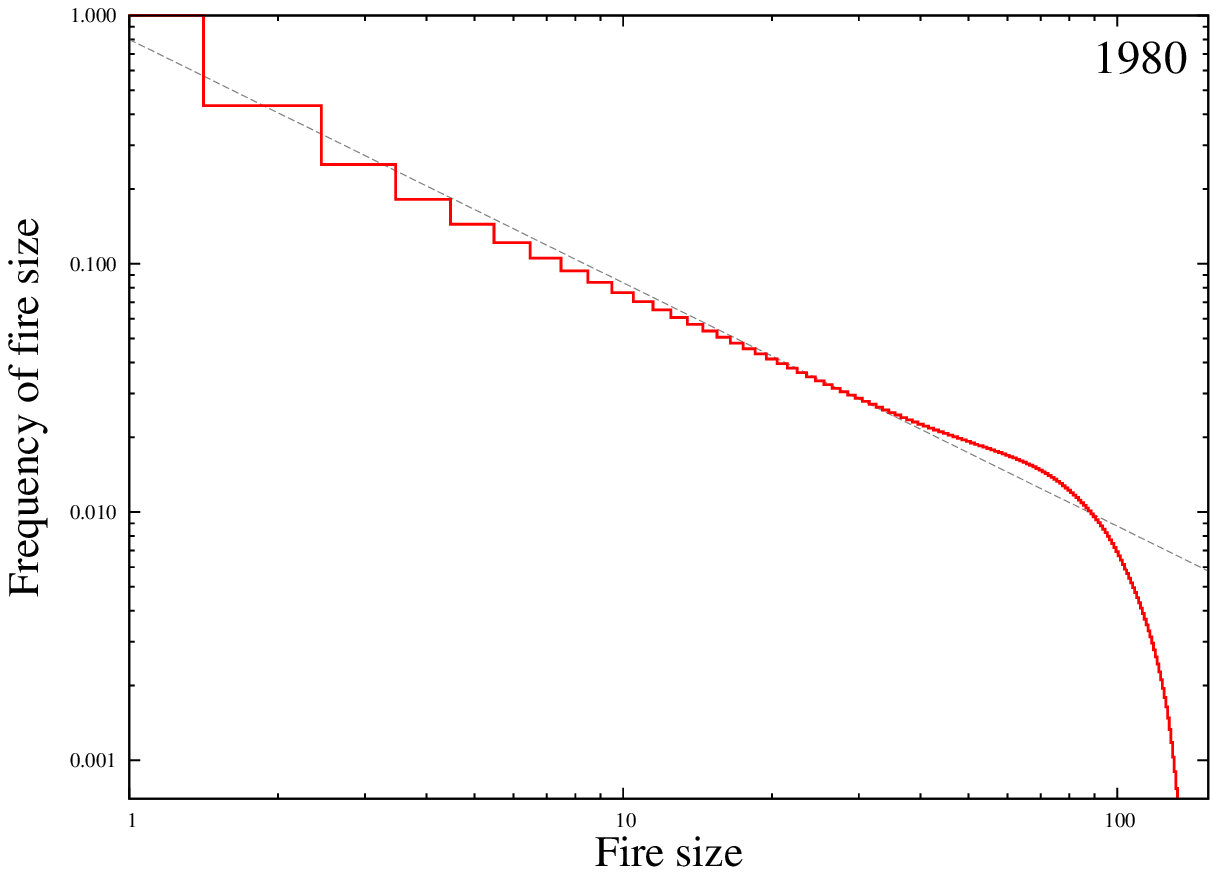}}
    \\
      \mbox{\includegraphics*[width=0.47\textwidth]{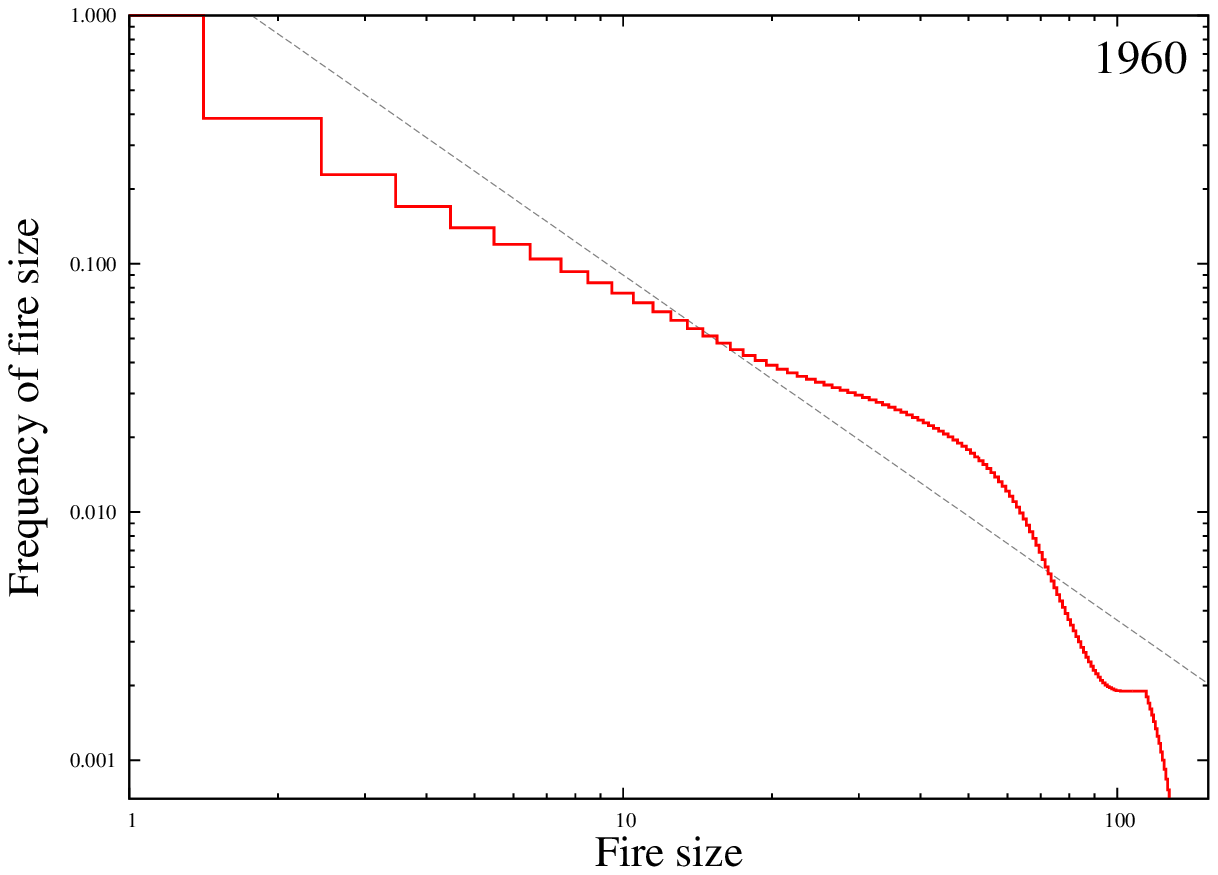}} &
      \mbox{\includegraphics*[width=0.47\textwidth]{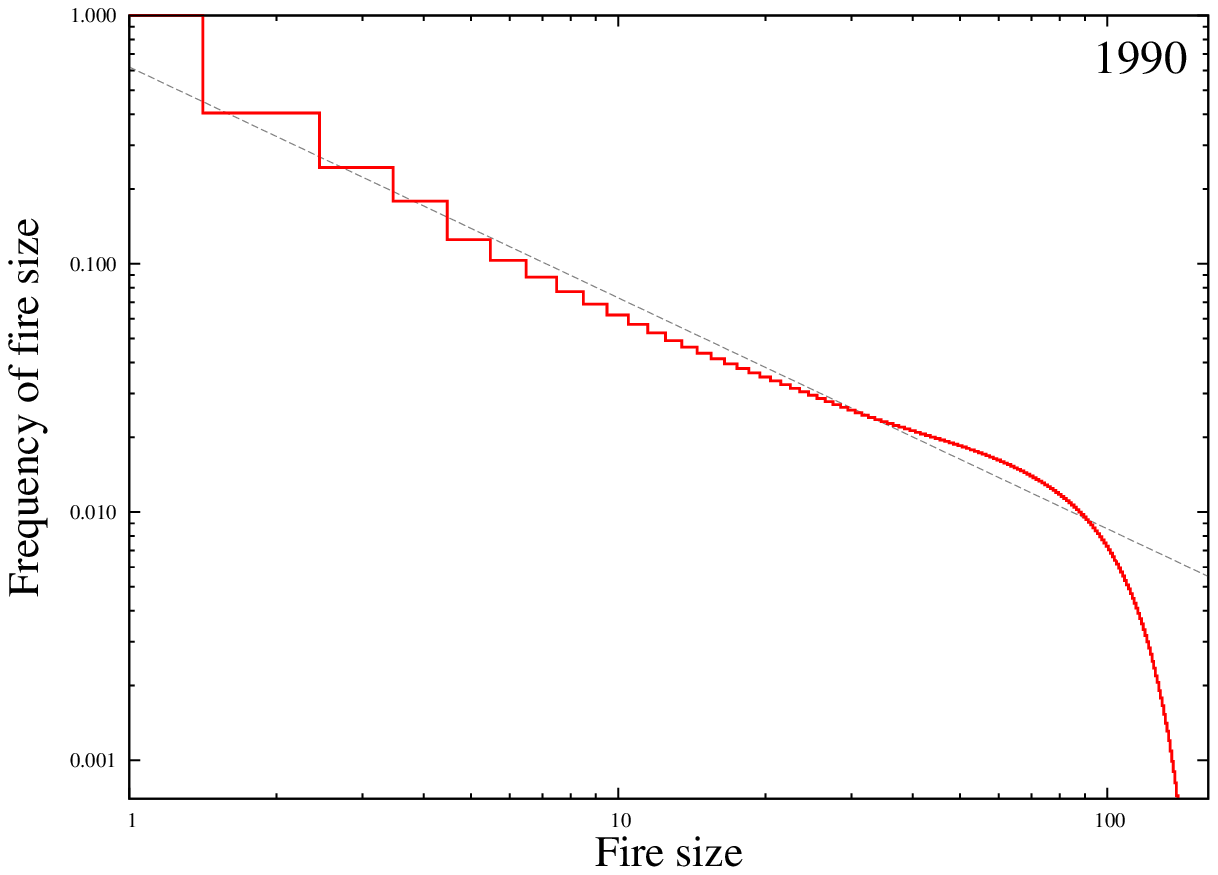}}
    \\
      \mbox{\includegraphics*[width=0.47\textwidth]{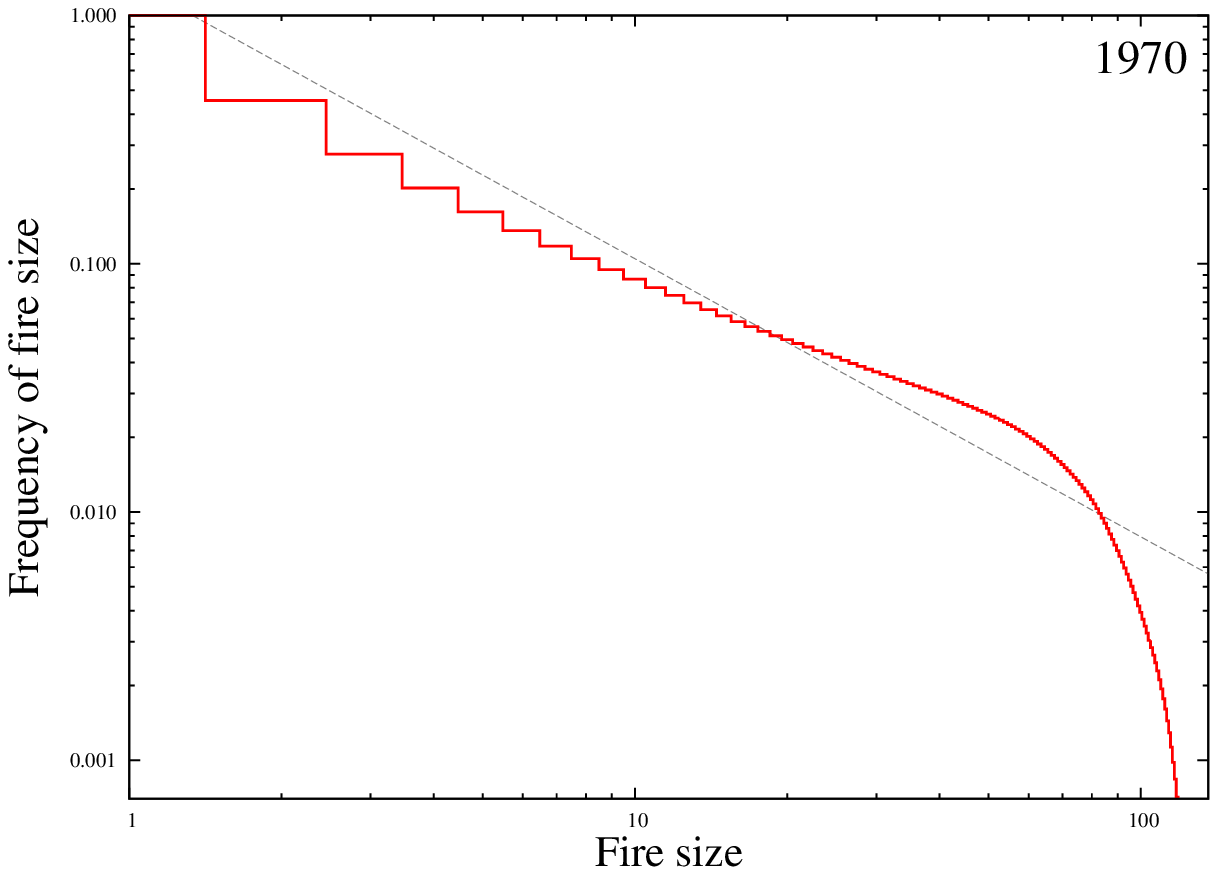}} &
      \mbox{\includegraphics*[width=0.47\textwidth]{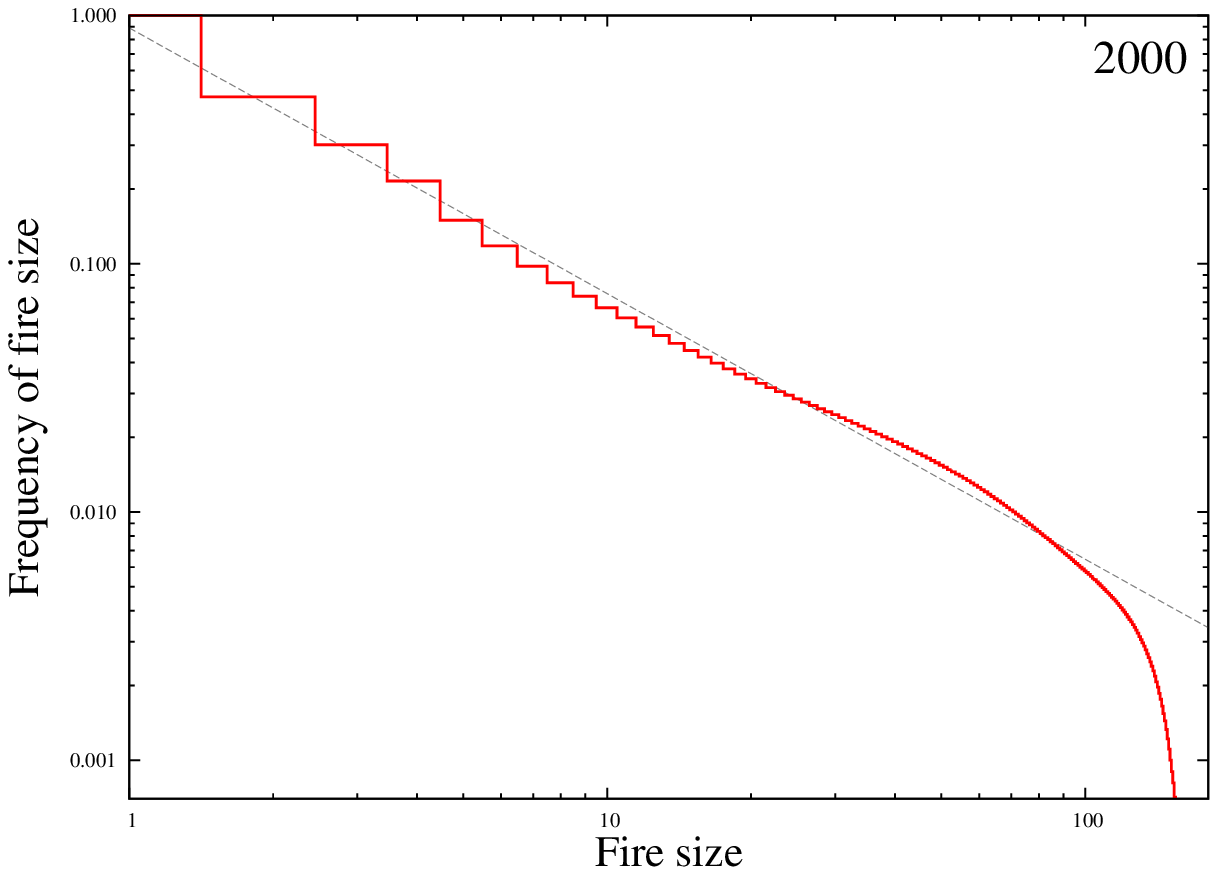}}
    \end{tabular}
  \end{center}
\caption{Frequency of firing size, i.e.\ the number of nodes involved in firing cascades, in log-log scale and for $10^3$~repetitions of the IFO dynamics. The straight lines are the best power-law fits to the data.}
  \label{fig:sizeh}
\end{figure}

Following the analysis, in Fig.~\ref{fig:sizeh} we present the probability distribution of sizes of cascades. These plots add more information to the previous discussion. As years go on, the WTW follows a cascade distribution that tends to a power law intermediate regime plus the obvious cutoff due to the finite size of the system. The power law structure of the distribution starts to be significant from 1970. The consequence of this power law is the absence of a characteristic scale of sizes of cascades in the system and implies that, in principle, the assessment of the scope of international crisis becomes unpredictable.

\begin{figure}[p]
  \begin{center}
    \begin{tabular}[t]{cc}
      \mbox{\includegraphics*[width=0.47\textwidth]{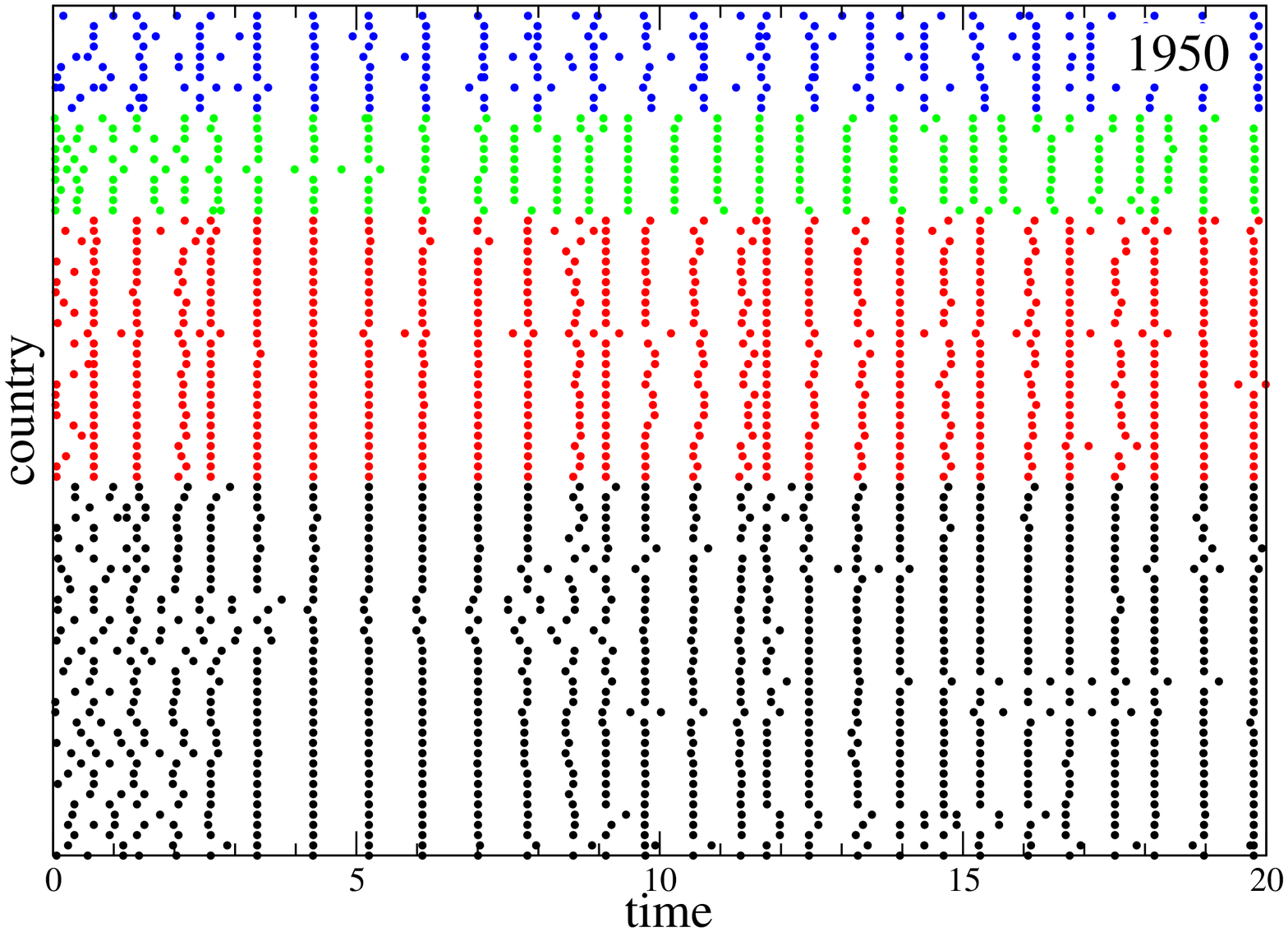}} &
      \mbox{\includegraphics*[width=0.47\textwidth]{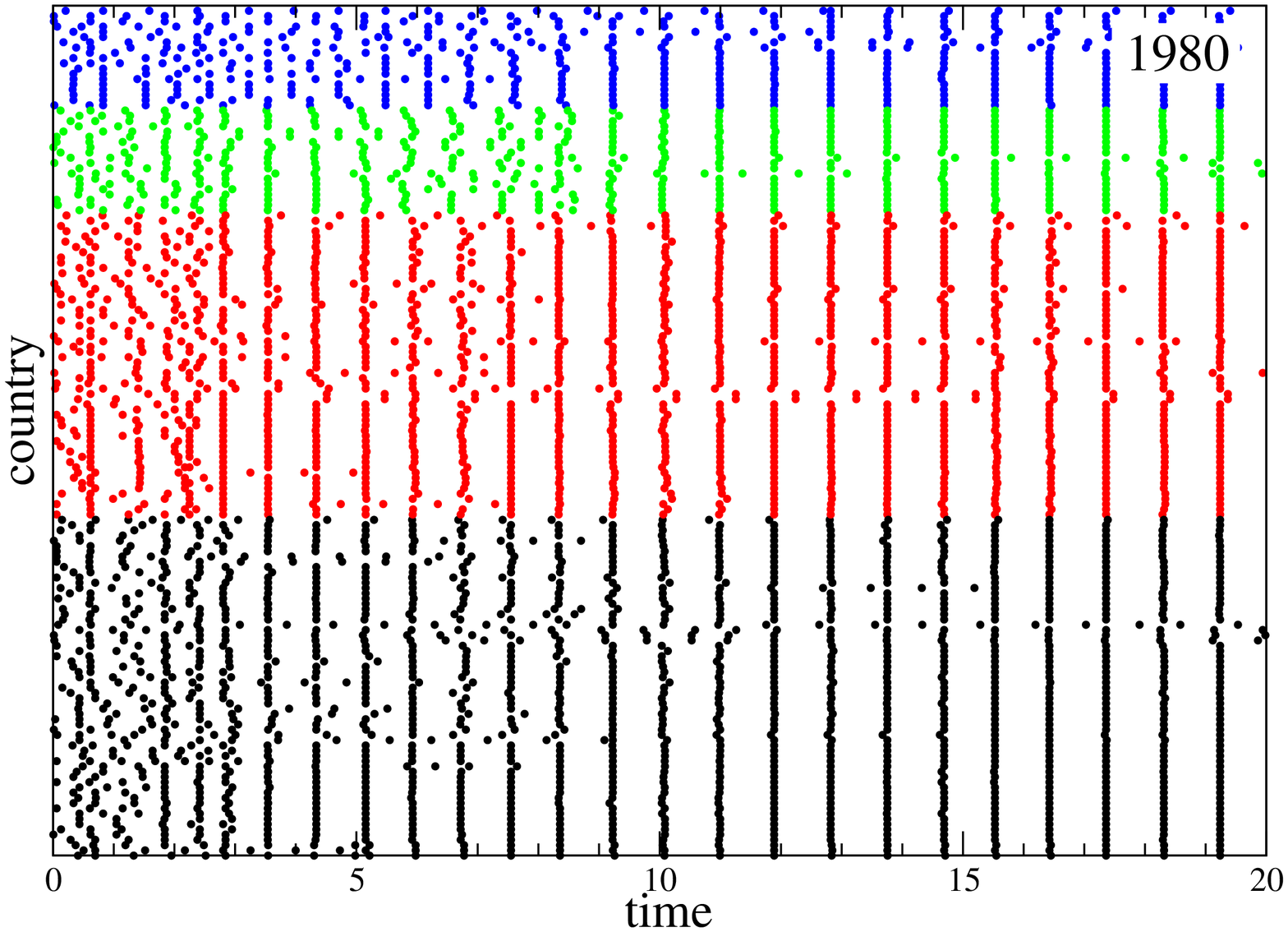}}
    \\
      \mbox{\includegraphics*[width=0.47\textwidth]{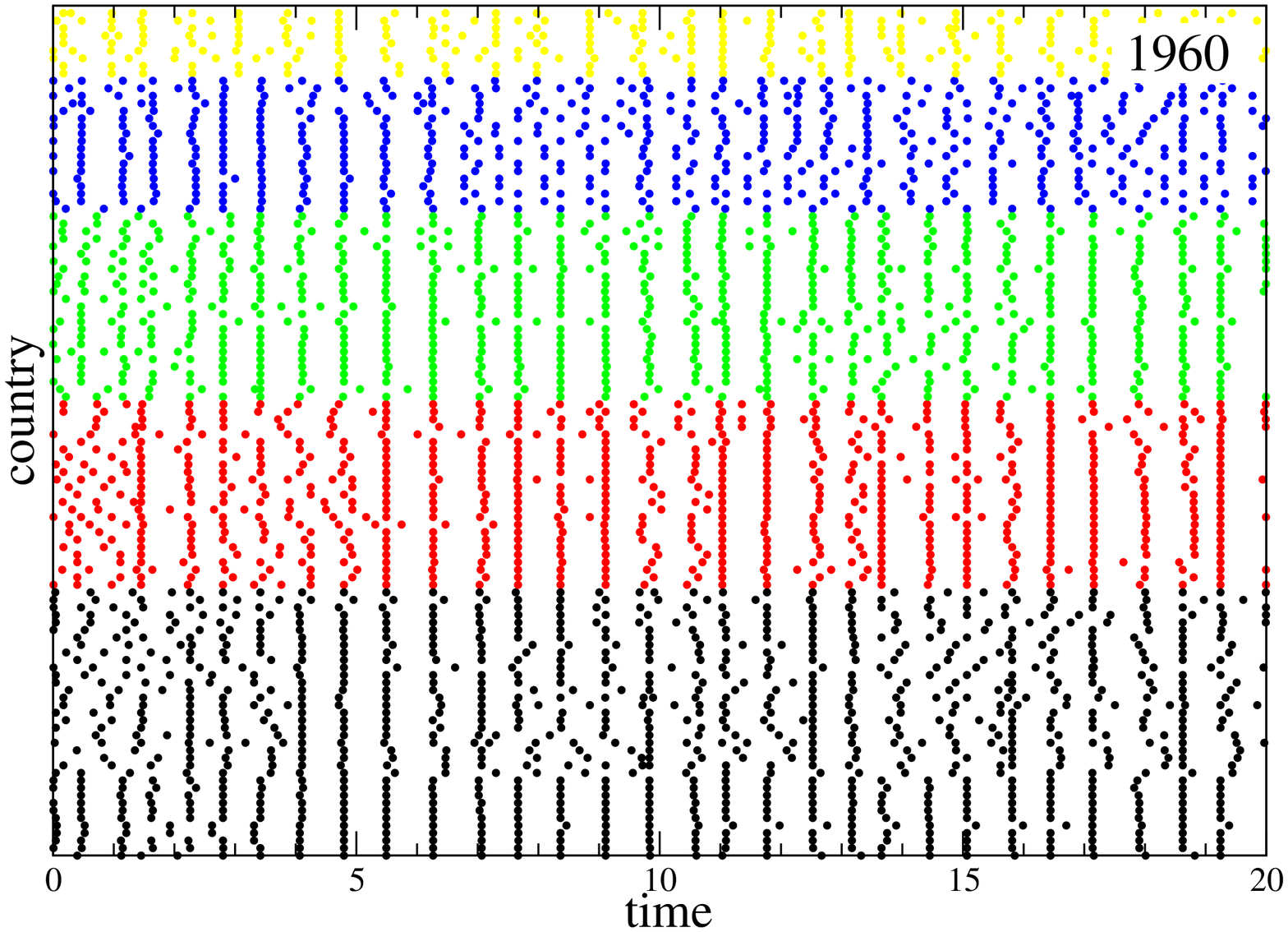}} &
      \mbox{\includegraphics*[width=0.47\textwidth]{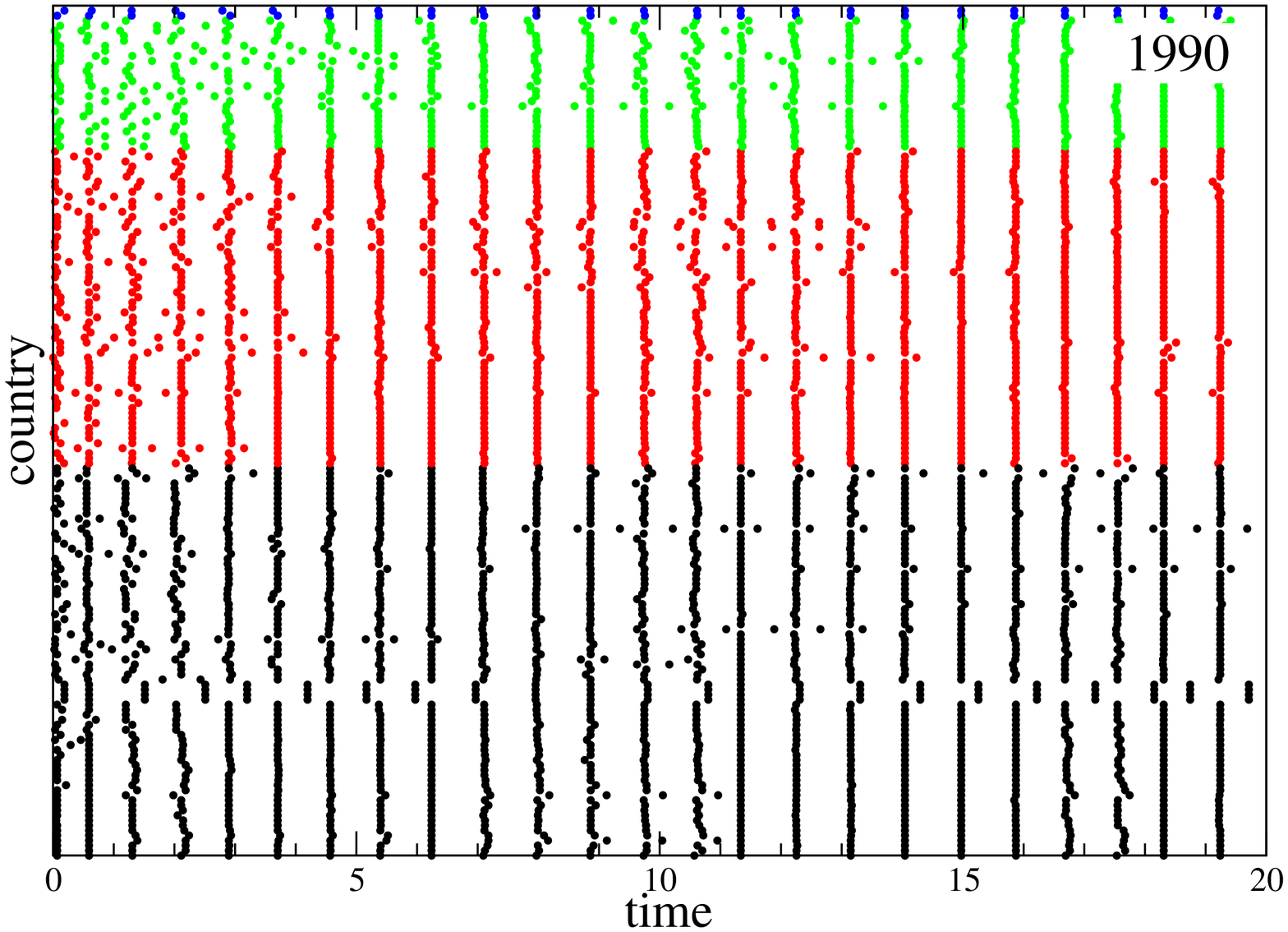}}
    \\
      \mbox{\includegraphics*[width=0.47\textwidth]{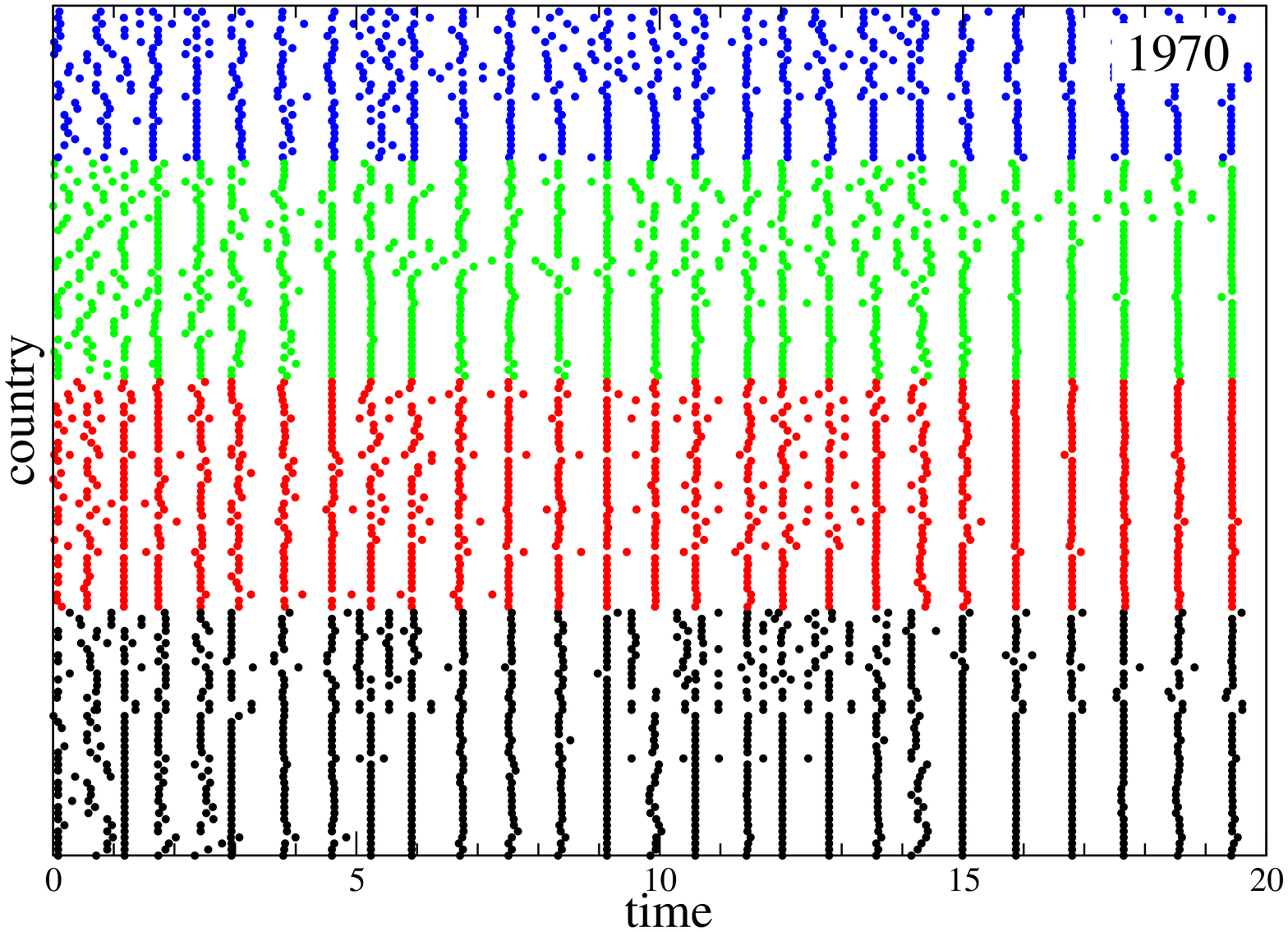}} &
      \mbox{\includegraphics*[width=0.47\textwidth]{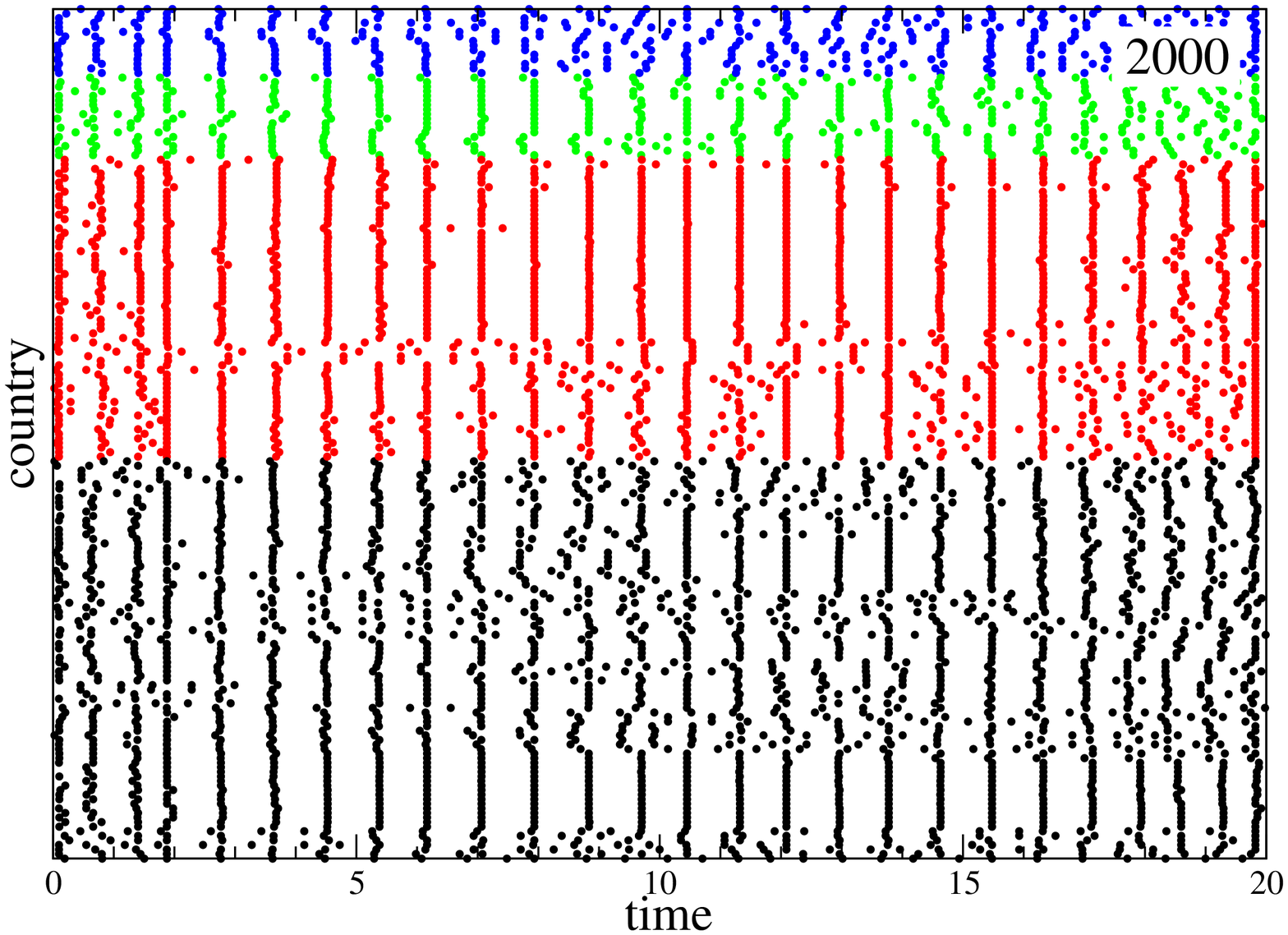}}
    \end{tabular}
  \end{center}
\caption{Firing times of the nodes in the first 20~time cycles of a single run of the IFO dynamics. Colors correspond to the communities in Fig.~\ref{fig:comm1}.}
  \label{fig:fire}
\end{figure}

\begin{figure}[p]
  \begin{center}
    \begin{tabular}[t]{cc}
      \mbox{\includegraphics*[width=0.47\textwidth]{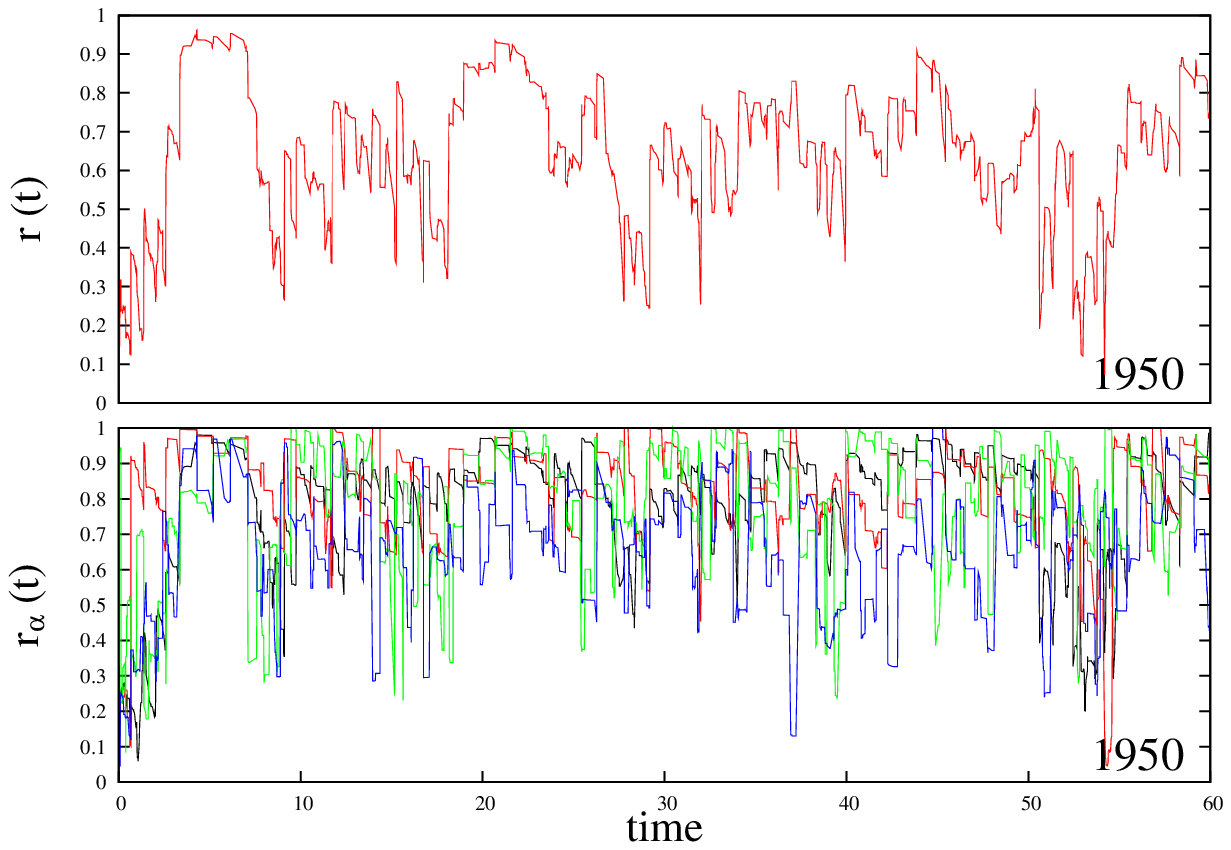}} &
      \mbox{\includegraphics*[width=0.47\textwidth]{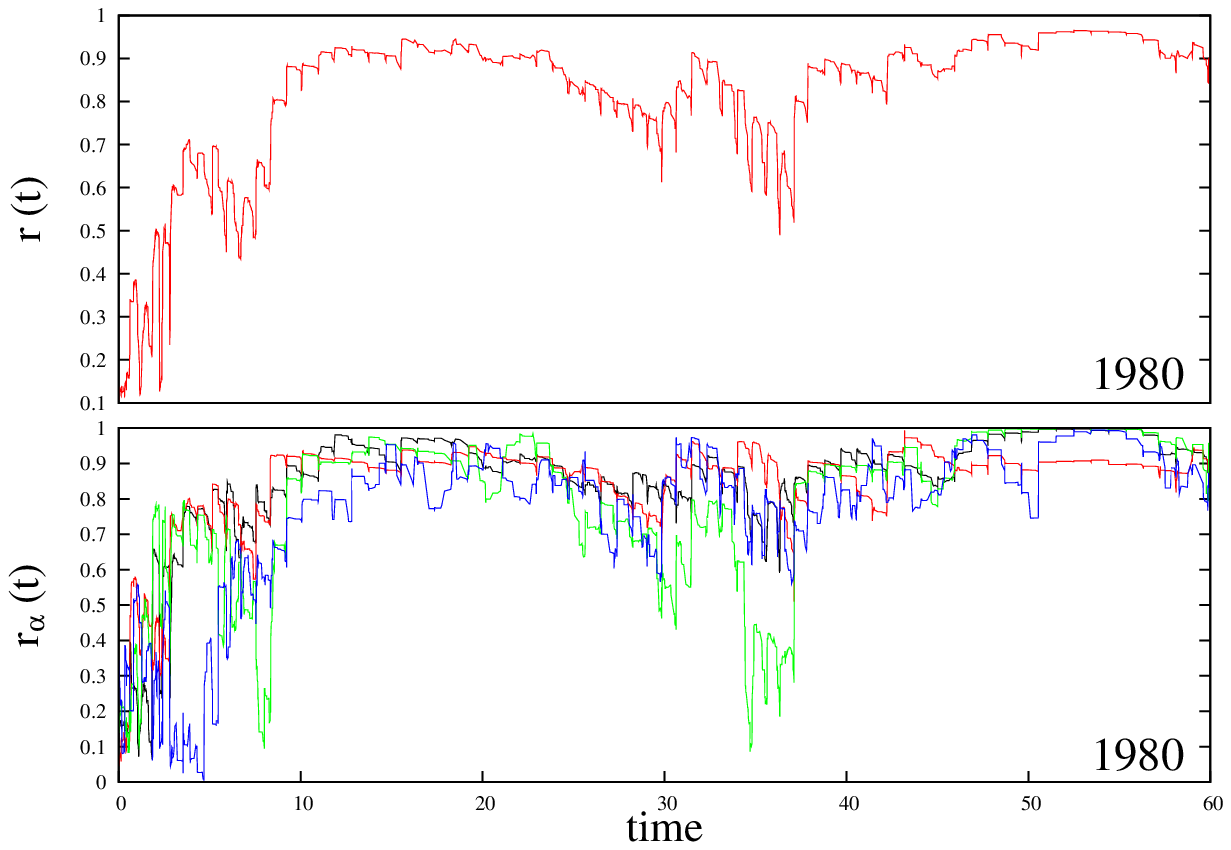}}
    \\
      \mbox{\includegraphics*[width=0.47\textwidth]{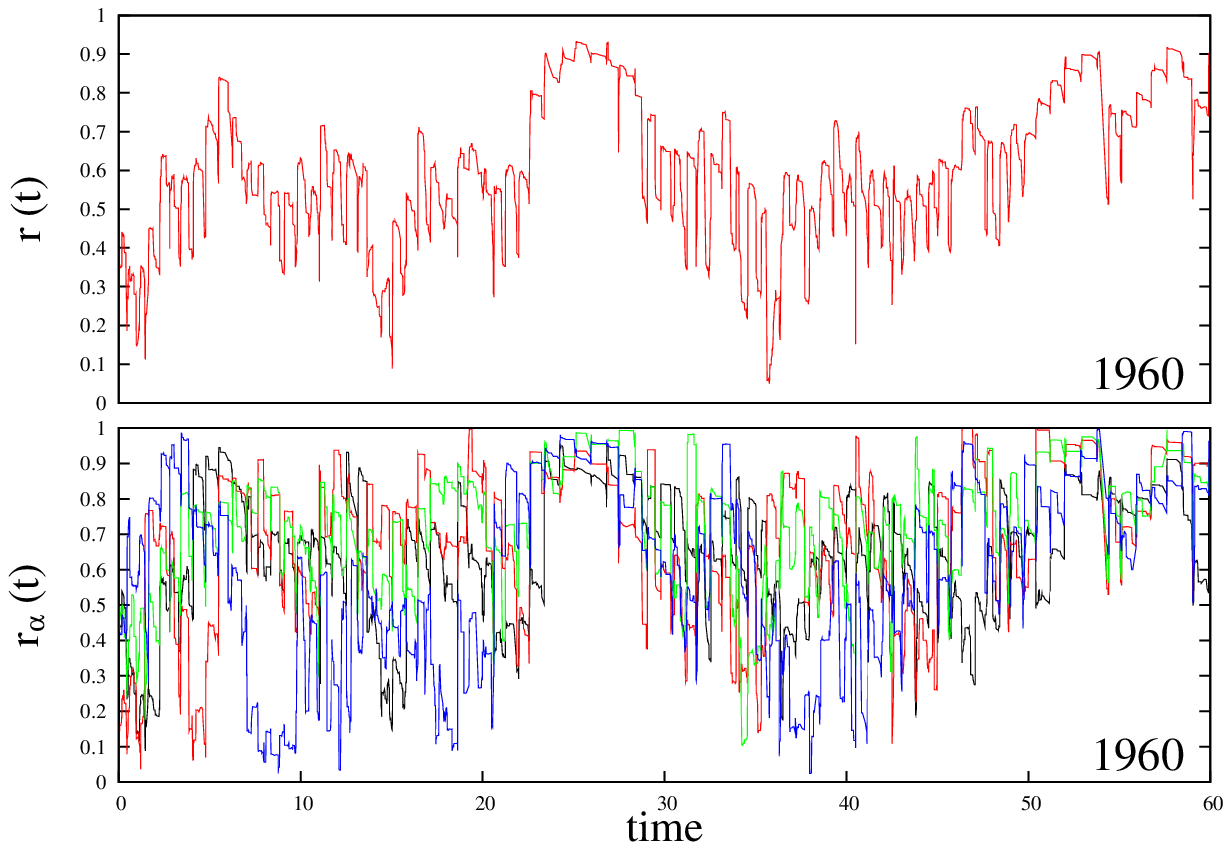}} &
      \mbox{\includegraphics*[width=0.47\textwidth]{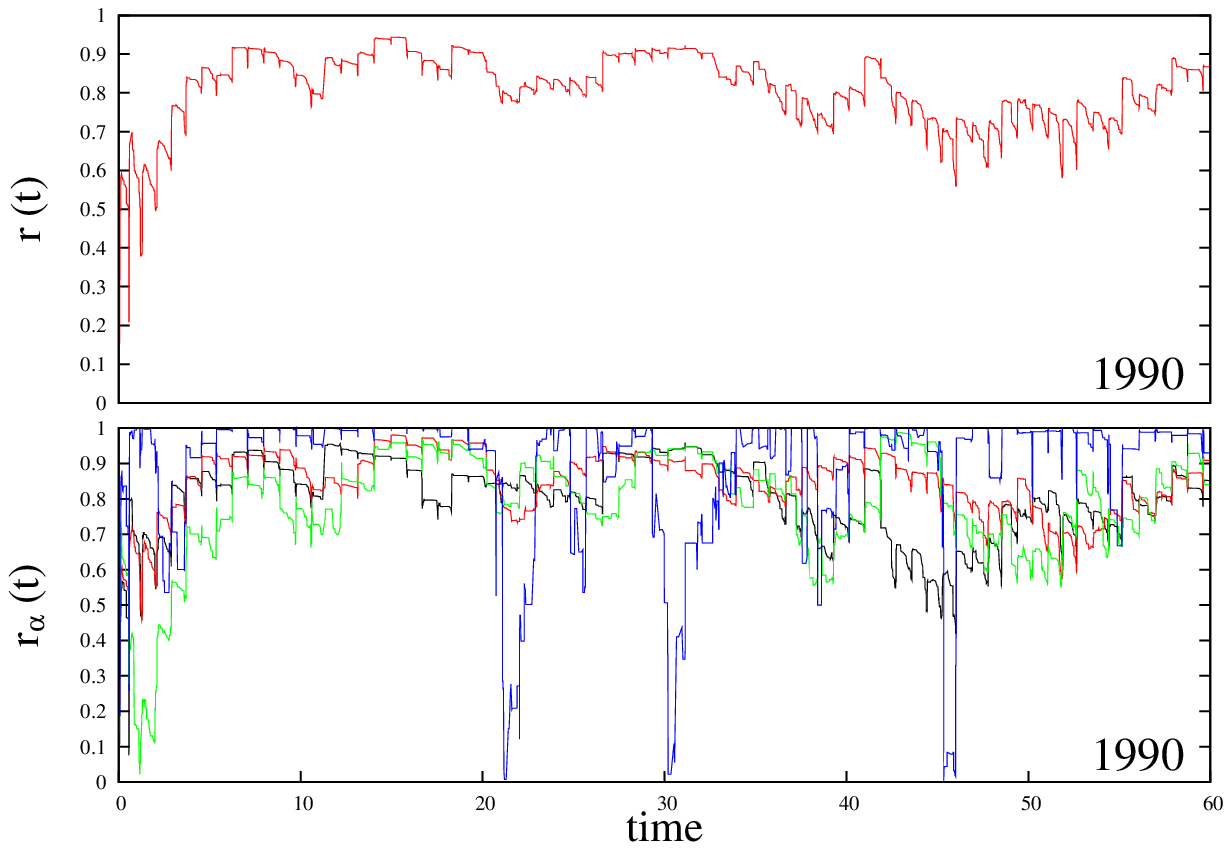}}
    \\
      \mbox{\includegraphics*[width=0.47\textwidth]{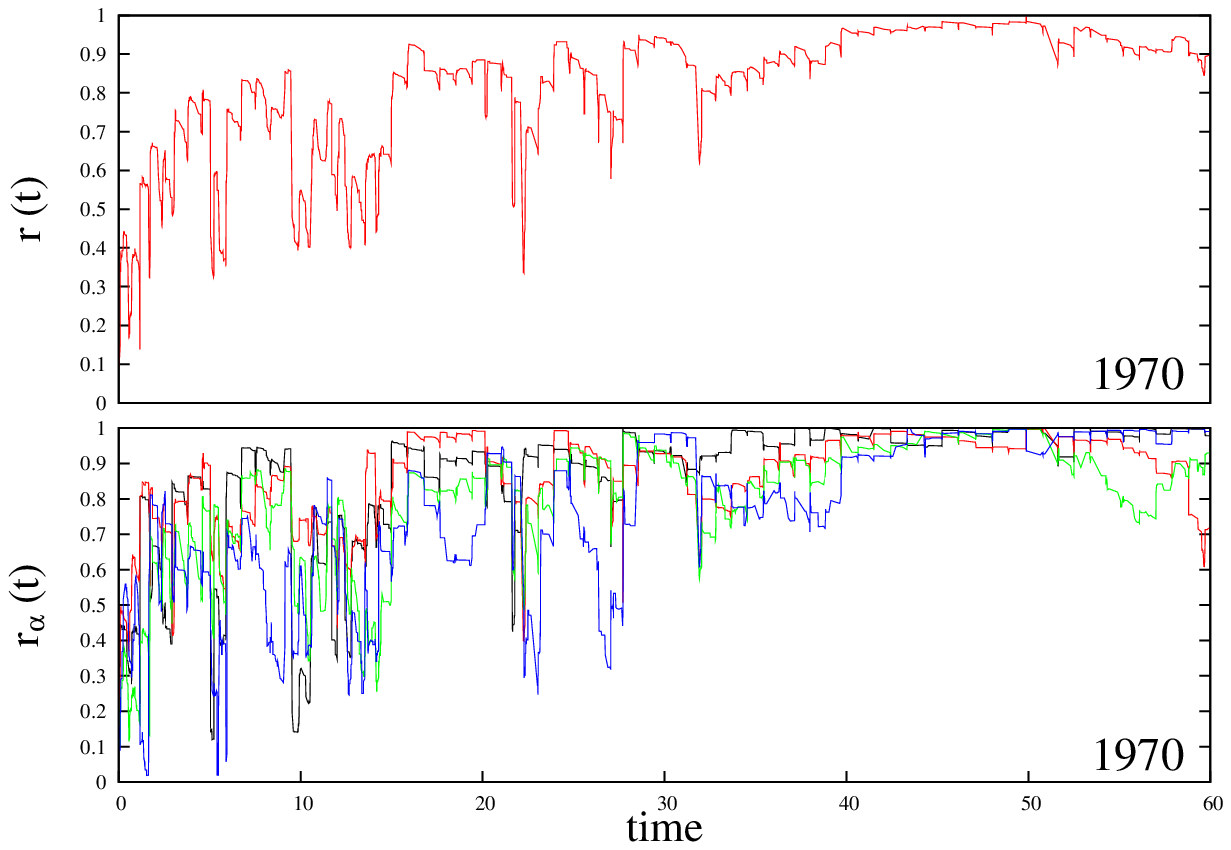}} &
      \mbox{\includegraphics*[width=0.47\textwidth]{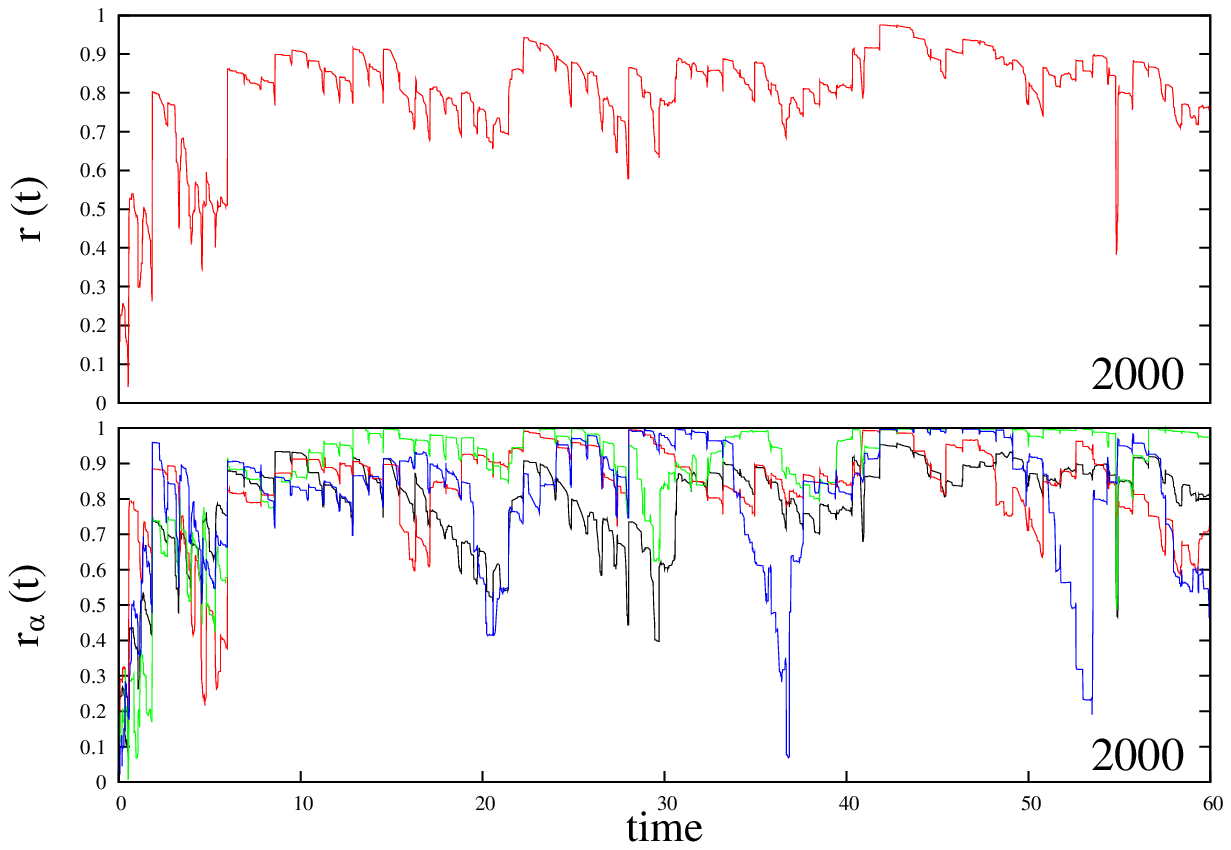}}
    \end{tabular}
  \end{center}
\caption{Evolution of the global and communities synchronization parameters, $r$ and $r_{\alpha}$ respectively, in the first 60~time cycles of a single run of the IFO dynamics. Colors in $r_{\alpha}$ plots correspond to the communities in Fig.~\ref{fig:comm1}.}
  \label{fig:rtt}
\end{figure}

Having analyzed the global outcome of the IFO dynamics on the WTW, we pay attention to the development of cascades restricted to the topological communities found. In Fig.~\ref{fig:fire} we show a raster plot of firings for each country, labeled with a color representing its topological community. Qualitatively, we detect that the communities behave more independently for the years 1950--1960--1970, a pre-globalization period, and more coherently between them for the 1980--1990--2000 periods. This is confirmed in Fig.~\ref{fig:rtt}, where we show the evolution of the synchronization parameter $r$ for each community and the whole network as a function of time. In this plot it can be observed that before and in 1970 (included) the synchronization parameter $r$ presents larger fluctuations than the global synchronization $r$. Moreover, it is clear that periods of synchronization and desynchronization of economical crisis coexist \cite{campbell1999} among different communities at the same time, and with the global measure. After 1970, the tendency is that of having a high level of synchronization in time with a very few desynchronization periods.

\begin{figure}[p]
  \begin{center}
    \begin{tabular}[t]{cc}
      \mbox{\includegraphics*[width=0.47\textwidth]{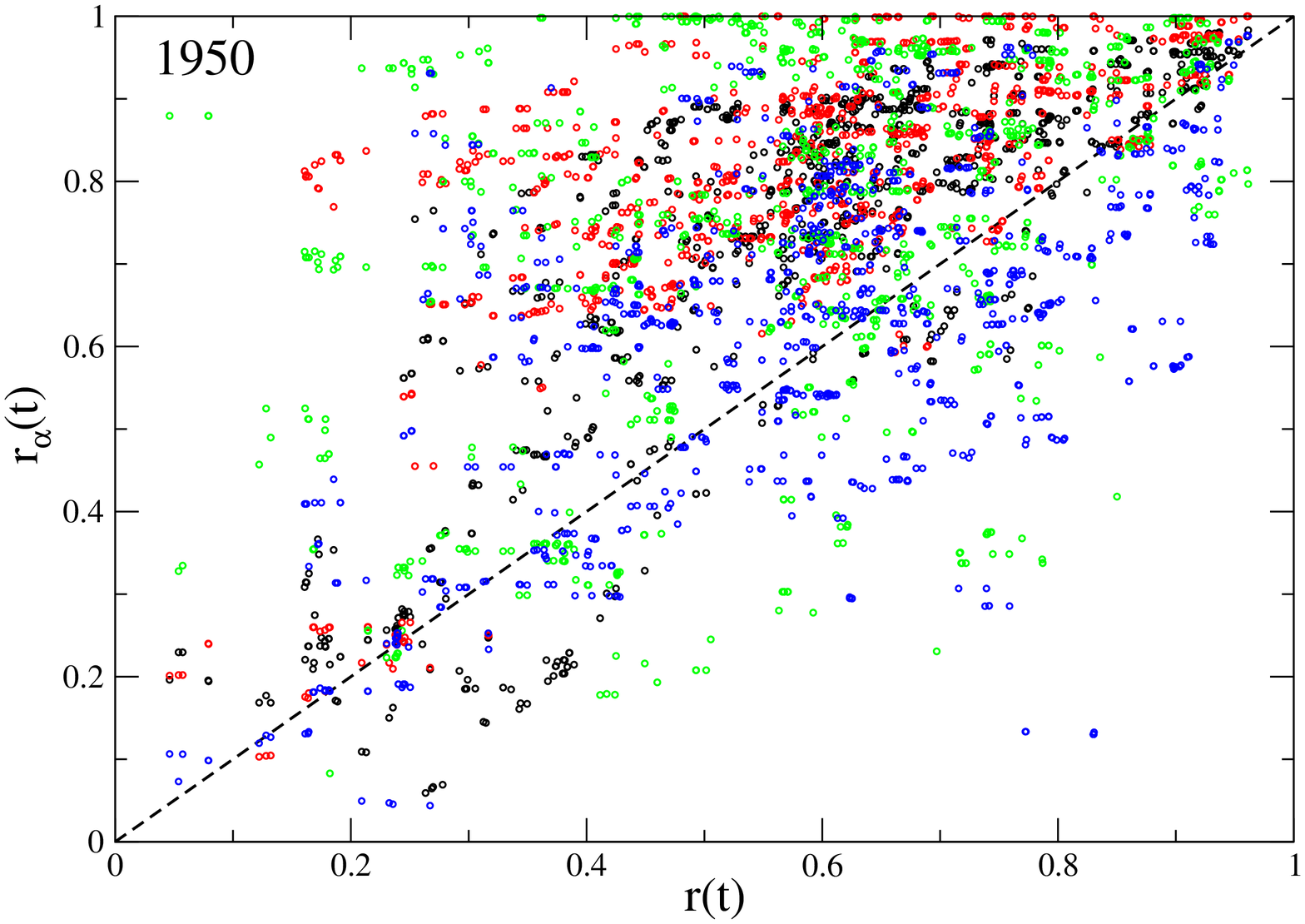}} &
      \mbox{\includegraphics*[width=0.47\textwidth]{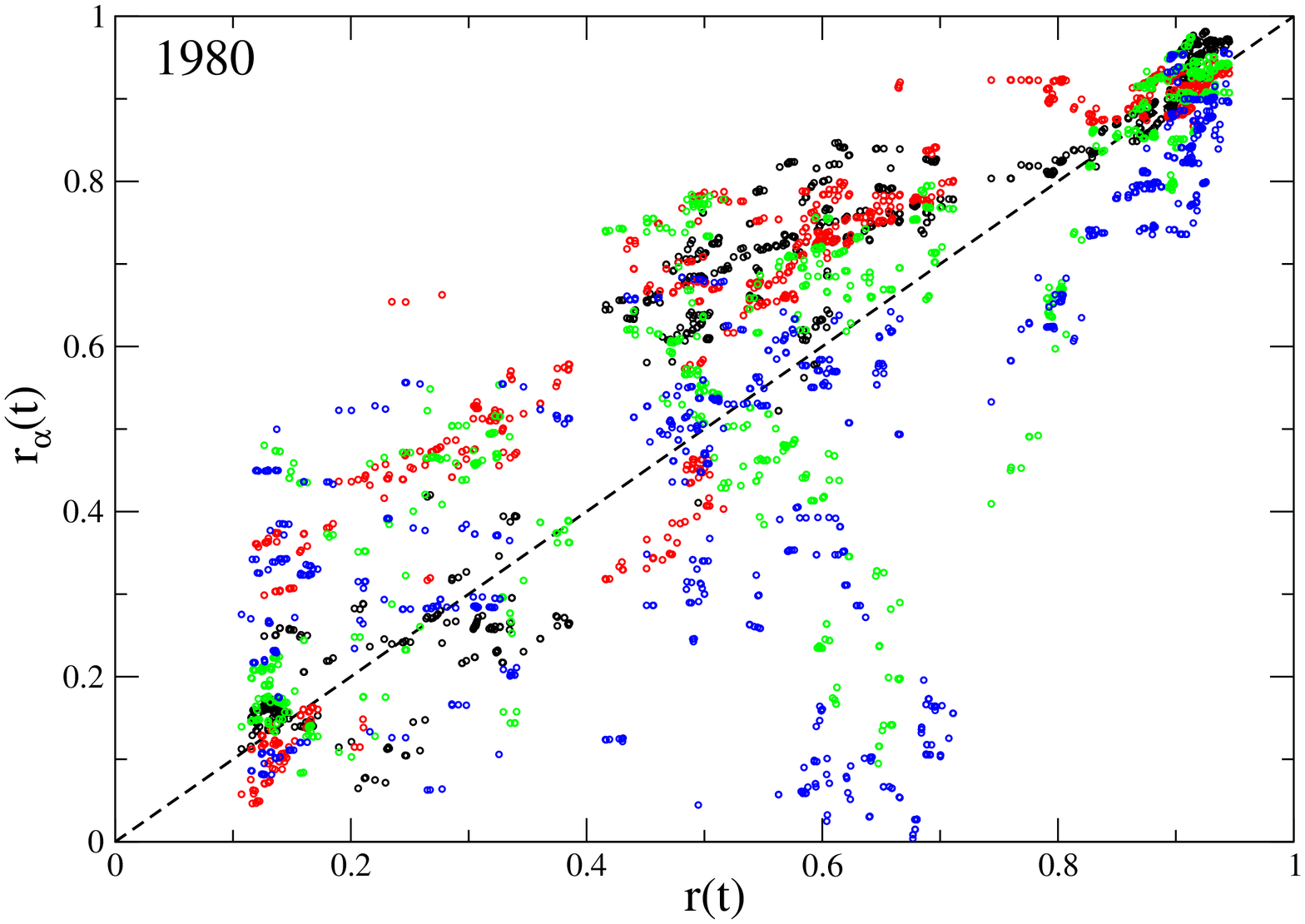}}
    \\
      \mbox{\includegraphics*[width=0.47\textwidth]{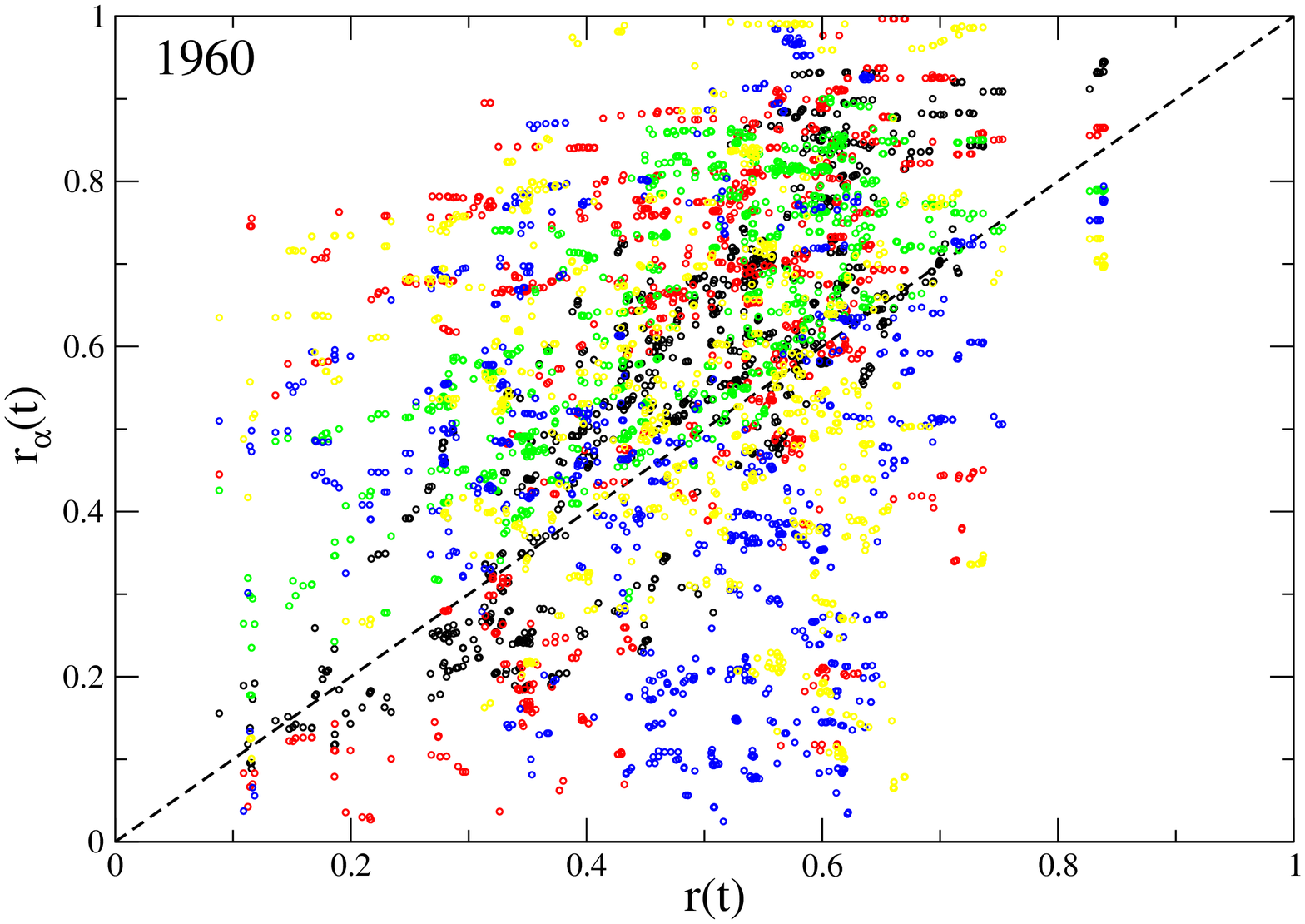}} &
      \mbox{\includegraphics*[width=0.47\textwidth]{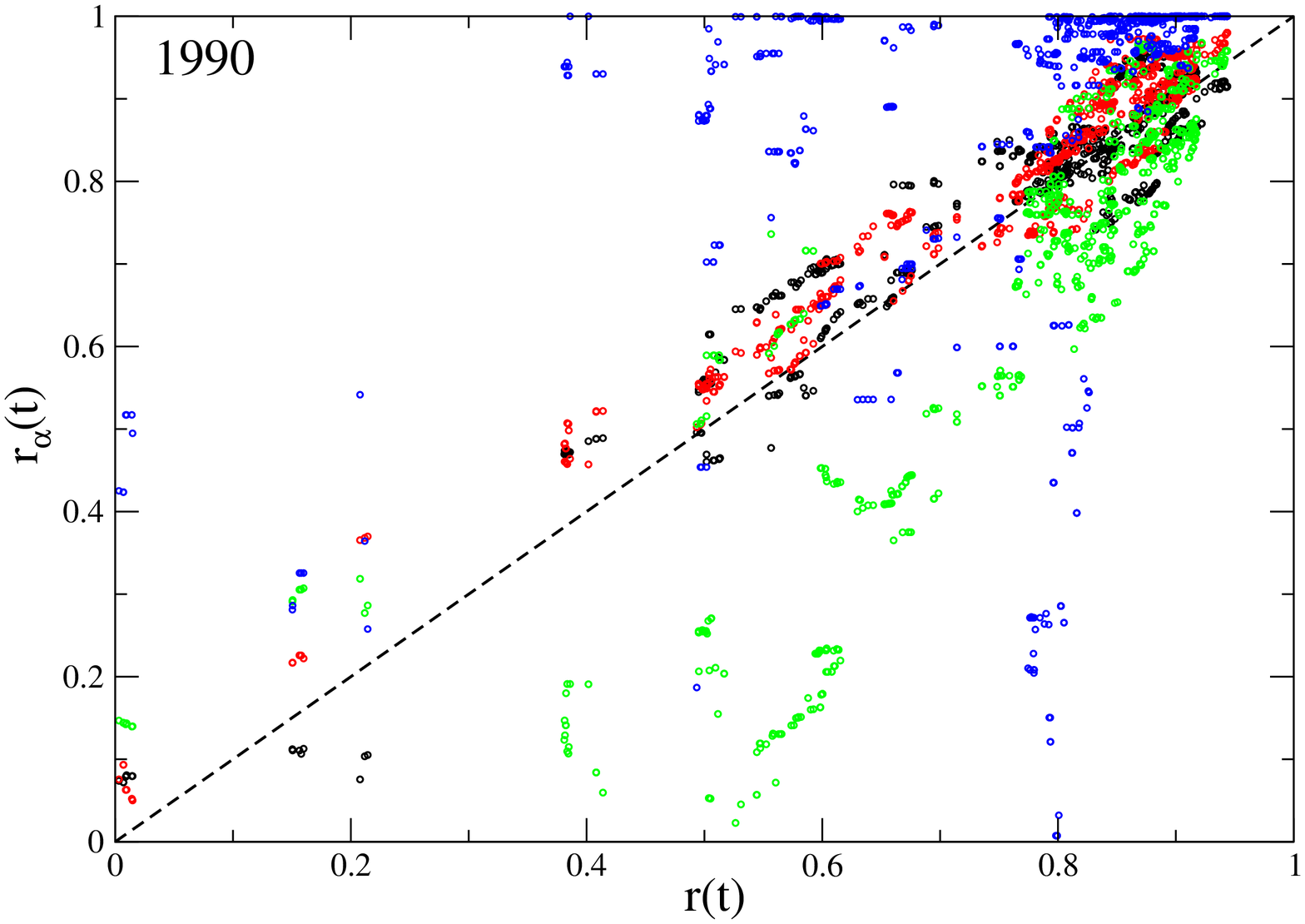}}
    \\
      \mbox{\includegraphics*[width=0.47\textwidth]{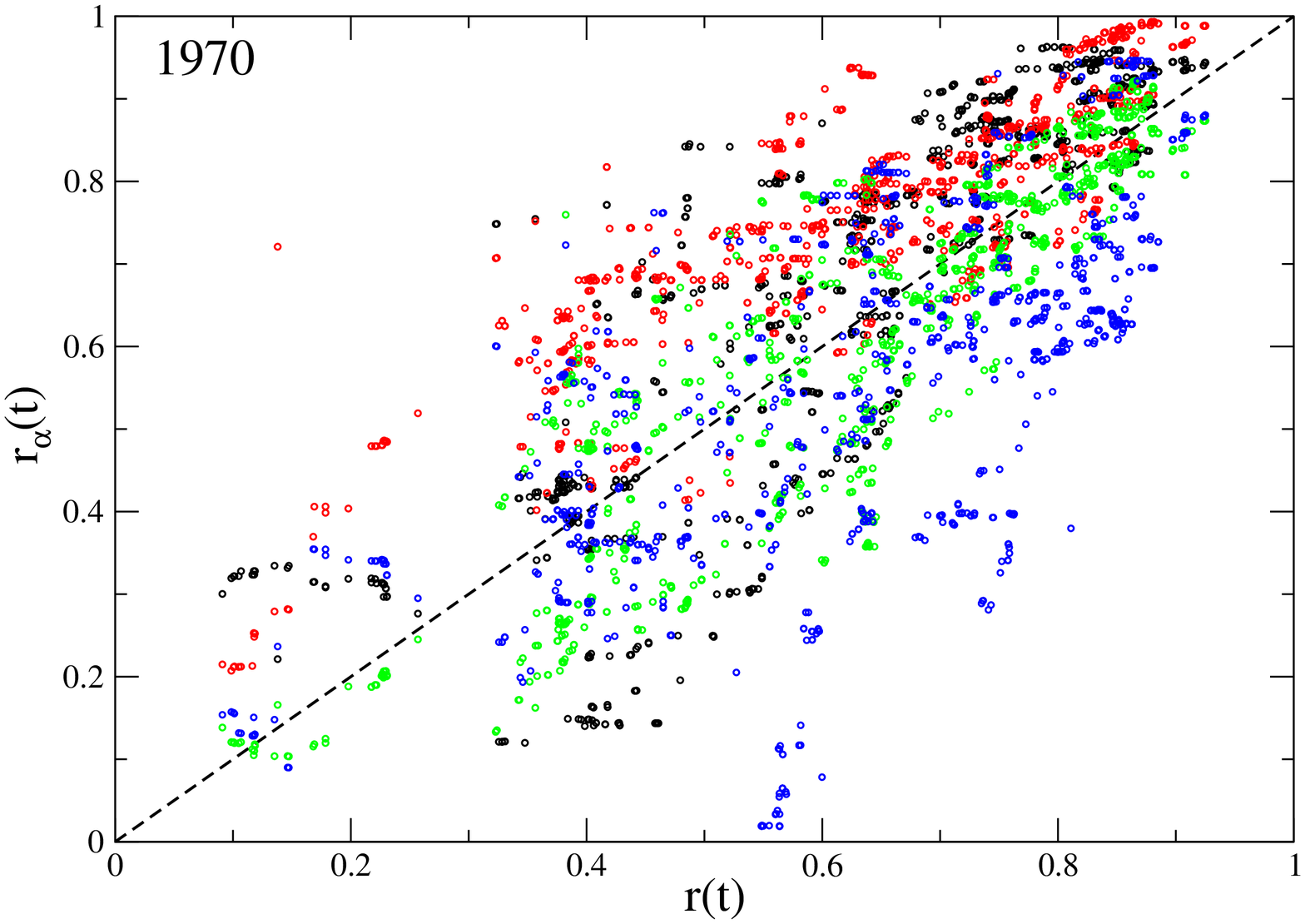}} &
      \mbox{\includegraphics*[width=0.47\textwidth]{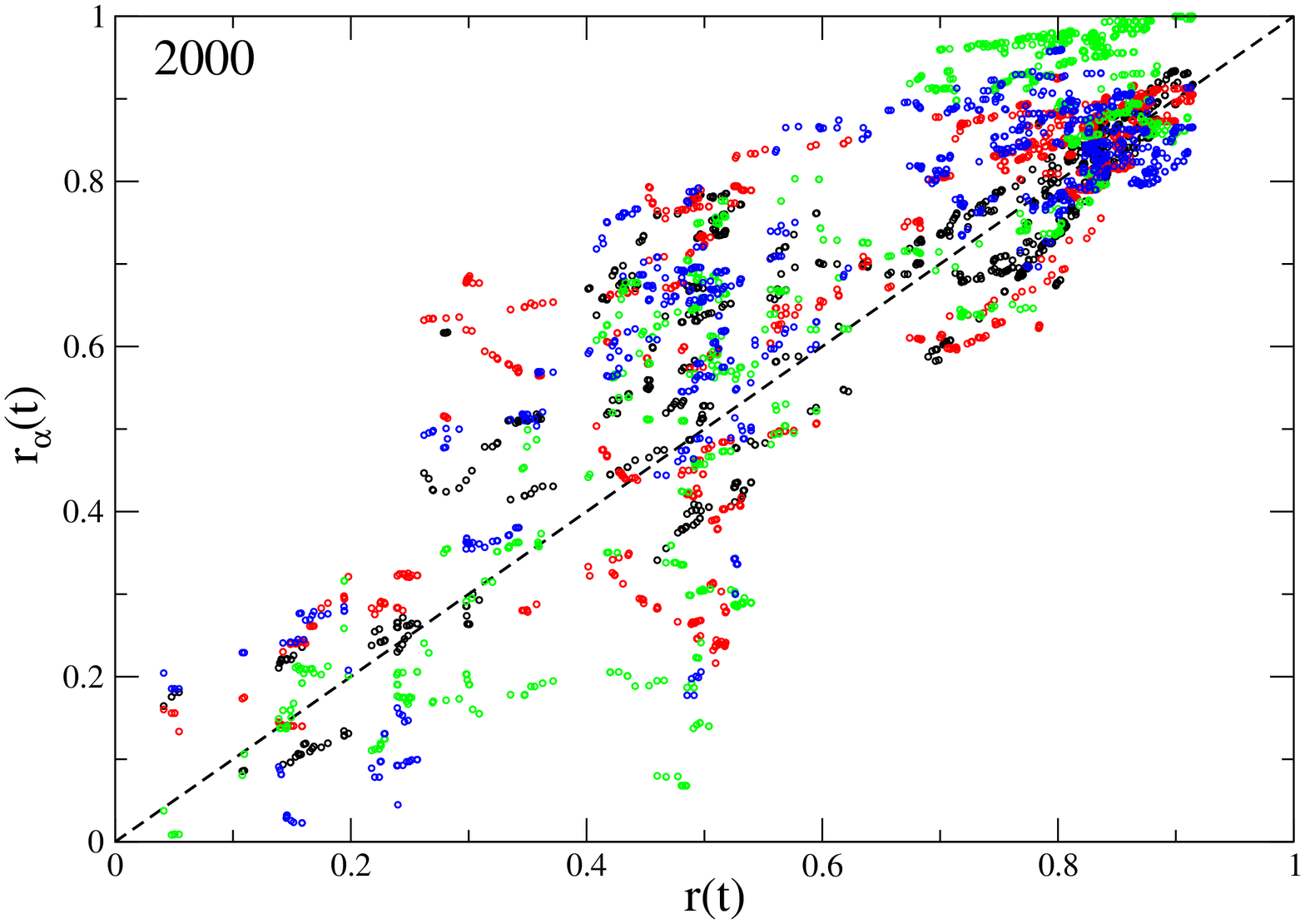}}
    \end{tabular}
  \end{center}
\caption{Deviation of the synchronization of the communities $r_{\alpha}$ in front of the global synchronization $r$, corresponding to the data in Fig.~\ref{fig:rtt}. Colors correspond to the communities in Fig.~\ref{fig:comm1}.}
  \label{fig:rart}
\end{figure}

Finally in Fig.~\ref{fig:rart}, we show the comparison of the values of $r$ of each community in front of that of all the network. The dispersion along the diagonal shows the discrepancies between the mesoscopic view and the global view. Again, after 1970 the signs of globalization are clear, the dynamic effect of the mesoscopic structure is practically collapsed to the global world scale behavior.

These observations allow us to conjecture that the effect of topological borders of the communities have almost no effect after globalization emerges.


\section{Conclusions}
In this paper we have presented a simple dynamical model of economic cycles interaction between countries in the WTW network. The model is represented by integrate and fire oscillators that emit pulses at the end of the economical cycles. We have proposed a longitudinal study to assess the effect of the mesoscopic structure of the WTW as time evolved. The results support the theory of a globalization process emerging in the decade 1970--1980, the synchronization phenomena after this period accelerates and the effect of a mesoscopic structure of  communities of countries is almost dissolved in the global behavior. The refining of the model presented via data driven approaches or by introducing more heterogeneity on the state of nodes can be a good way to investigate the scope of world crisis.


\section*{Acknowledgements}
This work has been supported by the Spanish MICINN FIS2009-13730-C02-01 and FIS2009-13730-C02-02, and the Generalitat de Catalunya 2009-SGR-838. PE acknowledges a URV PhD grant.


\medskip

\medskip

\appendix
\section{Community structure of the WTW}

\subsection*{1950}
\begin{description}
\item[Black] United Kingdom, Ireland, Netherlands, Belgium, Luxembourg, France, Switzerland, Spain, Portugal, French Guiana, Austria, Italy, Serbia, Greece, Finland, Sweden, Norway, Denmark, Germany, Iceland, South Africa, Iran, Turkey, Iraq, Egypt, Syria, Lebanon, Israel, Saudi Arabia, Mongolia, India, Bhutan, Pakistan, Sri Lanka, Nepal, Indonesia, Australia, New Zealand
\item[Red] United States, Canada, Cuba, Haiti, Dominican Republic, Mexico, Guatemala, Honduras, Santa Lucia, Nicaragua, Costa Rica, Panama, Colombia, Venezuela, Ecuador, Peru, Brazil, Bolivia, Paraguay, Chile, Argentina, Uruguay, Ethiopia, Yemen, Thailand, Philippines
\item[Green] Poland, Hungary, Albania, Bulgaria, Romania, Russia, China, South Korea
\item[Blue] Liechtenstein, Liberia, Jordan, Oman, Afghanistan, Taiwan, North Korea, Japan, Myanmar
\end{description}

\subsection*{1960}
\begin{description}
\item[Black] Paraguay, Argentina, Netherlands, Belgium, Luxembourg, France, Switzerland, Portugal, French Guiana, Austria, Italy, Mali, Senegal, Benin, Mauritania, Niger, Cote Ivoire, Togo, Cambodia, Central African Republic, Chad, Congo Republic, Somalia, Morocco, Tunisia, Turkey, Iraq, Syria, Lebanon, Saudi Arabia, Yemen
\item[Red] United States, Canada, Cuba, Haiti, Dominican Republic, Mexico, Guatemala, Honduras, Santa Lucia, Nicaragua, Costa Rica, Panama, Colombia, Venezuela, Ecuador, Peru, Brazil, Bolivia, Chile, Liberia, Ethiopia, Bhutan, Nepal, Philippines, Indonesia
\item[Green] United Kingdom, Ireland, Spain, Poland, Cyprus, Finland, Sweden, Norway, Denmark, Germany, Iceland, Ghana, Nigeria, South Africa, Sudan, Iran, Israel, Oman, India, Pakistan, Myanmar, Sri Lanka, Laos, Australia, New Zealand
\item[Blue] Uruguay, Hungary, Albania, Serbia, Greece, Bulgaria, Romania, Russia, Guinea, Egypt, Afghanistan, China, Mongolia, South Korea, Cameroon
\item[Yellow] Liechtenstein, Libya, Jordan, Taiwan, North Korea, Japan, Thailand, Malaysia
\end{description}

\subsection*{1970}
\begin{description}
\item[Black] United States, Canada, Haiti, Dominican Republic, Jamaica, Trinidad and Tobago, Mexico, Guatemala, Honduras, Santa Lucia, Nicaragua, Costa Rica, Panama, Colombia, Venezuela, Guyana, Ecuador, Peru, Brazil, Bolivia, Paraguay, Chile, Argentina, Uruguay, Liechtenstein, Rwanda, Ethiopia, Taiwan, North Korea, Japan, Bhutan, Thailand, Laos, Malaysia, Singapore, Philippines, Indonesia, Australia, Palestinian Authonomy
\item[Red] Netherlands, Belgium, Luxembourg, France, Switzerland, Spain, French Guiana, Austria, Italy, Mali, Senegal, Benin, Mauritania, Niger, Cote Ivoire, Liberia, Togo, Cambodia, Central African Republic, Chad, Congo Republic, Somalia, Morocco, Algeria, Tunisia, Libya, Iran, Turkey, Iraq, Saudi Arabia, Kuwait, Oman, Madagascar
\item[Green] Barbados, United Kingdom, Ireland, Portugal, Malta, Greece, Cyprus, Finland, Sweden, Norway, Denmark, Germany, Iceland, Gambia, Guinea, Sierra Leone, Ghana, Nigeria, Uganda, Kenya, Tanzania, Burkina, Zambia, Malawi, South Africa, Botswana, Swaziland, Mauritius, Israel, Yemen, Myanmar, Sri Lanka, New Zealand, Fiji
\item[Blue] Cuba, Poland, Hungary, Albania, Serbia, Bulgaria, Romania, Russia, Lesotho, Sudan, Egypt, Syria, Lebanon, Jordan, Afghanistan, China, Mongolia, South Korea, India, Pakistan, Nepal, Cameroon
\end{description}

\subsection*{1980}
\begin{description}
\item[Black] Dominica, Greenland, St Vincent and Grenadines, United Kingdom, Ireland, Netherlands, Belgium, Luxembourg, France, Switzerland, Spain, Portugal, French Guiana, Austria, Italy, Malta, Greece, Finland, Sweden, Norway, Denmark, Germany, Iceland, Sao Tome and Principe, Guinea Bissau, Mali, Senegal, Mauritania, Niger, Cote Ivoire, Guinea, Liberia, Sierra Leone, Togo, Cambodia, Central African Republic, Chad, Congo Republic, Uganda, Kenya, Tanzania, Burkina, Somalia, Zambia, Malawi, South Africa, Swaziland, Comoros, Mauritius, Morocco, Tunisia, Egypt, Syria, Lebanon, Israel, Saudi Arabia, Qatar, United Arab Emirates, Madagascar
\item[Red] United States, Canada, Haiti, Dominican Republic, Jamaica, Trinidad and Tobago, Barbados, Mexico, Guatemala, Honduras, Santa Lucia, Nicaragua, Costa Rica, Panama, Colombia, Venezuela, Guyana, Suriname, Ecuador, Peru, Liechtenstein, Gambia, Benin, Nigeria, Rwanda, Angola, Botswana, Seychelles, Algeria, Libya, Yemen, Bahamas, Oman, China, Taiwan, North Korea, Japan, Bhutan, Myanmar, Thailand, Cameroon, Malaysia, Singapore, Philippines, Indonesia, Australia, Papua New Guinea, New Zealand, Solomon Islands, Fiji, Palestinian Authonomy
\item[Green] Brazil, Bolivia, Paraguay, Chile, Argentina, Uruguay, Cyprus, Cabo Verde, Mozambique, Lesotho, Sudan, Iran, Turkey, Iraq, Jordan, Kuwait, Pakistan, Bangladesh, Sri Lanka
\item[Blue] Cuba, Poland, Hungary, Albania, Serbia, Bulgaria, Romania, Russia, Ghana, Ethiopia, Afghanistan, Mongolia, South Korea, India, Nepal, Laos
\end{description}

\subsection*{1990}
\begin{description}
\item[Black] United States, Canada, Haiti, Dominican Republic, Jamaica, Trinidad and Tobago, Barbados, Dominica, Greenland, St Vincent and Grenadines, St Kitts and Nevis, Mexico, Belize, Guatemala, Honduras, Santa Lucia, Costa Rica, Panama, Colombia, Venezuela, Guyana, Ecuador, Peru, Brazil, Bolivia, Paraguay, Chile, Argentina, Uruguay, Monaco, Liechtenstein, Benin, Liberia, Nigeria, Somalia, Angola, Mozambique, Lesotho, Sudan, Iraq, Jordan, Saudi Arabia, Yemen, Bahamas, Qatar, United Arab Emirates, Oman, China, Taiwan, North Korea, Japan, Bhutan, Pakistan, Bangladesh, Myanmar, Sri Lanka, Madagascar, Nepal, Thailand, Laos, Malaysia, Singapore, Brunei, Philippines, Indonesia, Australia, Papua New Guinea, New Zealand, Solomon Islands, Fiji, Palestinian Authonomy
\item[Red] Antigua and Barbuda, Suriname, United Kingdom, Ireland, Netherlands, Belgium, Luxembourg, France, Switzerland, Spain, Portugal, French Guiana, Austria, Italy, Malta, Greece, Finland, Sweden, Norway, Denmark, Germany, Iceland, Sao Tome and Principe, Guinea Bissau, Gambia, Mali, Senegal, Mauritania, Niger, Cote Ivoire, Guinea, Sierra Leone, Ghana, Cambodia, Central African Republic, Chad, Congo Republic, Kenya, Burkina, Rwanda, Zambia, Malawi, South Africa, Namibia, Botswana, Swaziland, Comoros, Mauritius, Seychelles, Morocco, Algeria, Tunisia, Libya, Iran, Turkey, Israel, Kuwait
\item[Green] Cuba, Nicaragua, Poland, Hungary, Albania, Serbia, Cyprus, Bulgaria, Romania, Russia, Cabo Verde, Uganda, Tanzania, Ethiopia, Egypt, Syria, Lebanon, Afghanistan, Mongolia, South Korea, India, Cameroon
\item[Blue] Togo, Burkina Faso
\end{description}

\subsection*{2000}
\begin{description}
\item[Black] Cuba, Barbados, Greenland, St Vincent and Grenadines, Antigua and Barbuda, United Kingdom, Ireland, Netherlands, Belgium, Luxembourg, France, Switzerland, Spain, Portugal, French Guiana, Poland, Austria, Hungary, Czech Republic, Slovakia, Italy, Malta, Albania, Macedonia, Croatia, Serbia, Bosnia and Herzegovina, Slovenia, Greece, Cyprus, Bulgaria, Maldives, Romania, Russia, Estonia, Latvia, Lithuania, Ukraine, Belarus, Armenia, Georgia, Azerbaijan, Finland, Sweden, Norway, Denmark, Germany, Iceland, Cabo Verde, Sao Tome and Principe, Mali, Mauritania, Niger, Cote Ivoire, Guinea, Sierra Leone, Cambodia, Central African Republic, Chad, Mozambique, Zambia, Malawi, South Africa, Namibia, Botswana, Swaziland, Comoros, Mauritius, Morocco, Algeria, Tunisia, Libya, Iran, Turkey, Syria, Turkmenistan, Tajikistan, Kyrgyzstan, Uzbekistan, Kazakhstan
\item[Red] United States, Canada, Haiti, Dominican Republic, Jamaica, Trinidad and Tobago, Dominica, St Kitts and Nevis, Mexico, Belize, Guatemala, Honduras, Santa Lucia, Nicaragua, Costa Rica, Panama, Colombia, Venezuela, Guyana, Suriname, Ecuador, Monaco, Liechtenstein, Benin, Liberia, Angola, Lesotho, Israel, Yemen, Kuwait, Qatar, United Arab Emirates, Oman, China, Mongolia, Taiwan, North Korea, Japan, India, Bhutan, Pakistan, Bangladesh, Myanmar, Sri Lanka, Madagascar, Nepal, Thailand, Cameroon, Laos, Malaysia, Singapore, Brunei, Philippines, Indonesia, Australia, Papua New Guinea, New Zealand, Solomon Islands, Fiji, Palestinian Authonomy
\item[Green] Uganda, Kenya, Tanzania, Burkina, Rwanda, Somalia, Ethiopia, Puerto Rico, Seychelles, Sudan, Iraq, Egypt, Lebanon, Jordan, Saudi Arabia, Bahamas, Afghanistan
\item[Blue] Peru, Brazil, Bolivia, Paraguay, Chile, Argentina, Uruguay, Guinea Bissau, Gambia, Senegal, Ghana, Togo, Nigeria, Congo Republic, South Korea
\end{description}

\newpage

\begin{table}[t]
  \begin{center}
    \begin{tabular}{c|*{6}{r}cl}
      \hline
      \multicolumn{1}{c|}{year} &  & \multicolumn{1}{c}{$N$} & \multicolumn{1}{c}{$D$} & \multicolumn{1}{c}{$\langle k \rangle$} & \multicolumn{1}{c}{$\langle s \rangle$} & \multicolumn{1}{c}{$Q_{\max}$} & \multicolumn{1}{c}{$M$} & \multicolumn{1}{c}{$N_{\alpha}$} \\
      \hline
      1950 & &  83 & 0.1753 &  28.75 &   705.99 & 0.4519 & 4 & [37, 26, 10, 10]    \\
      1960 & & 113 & 0.1670 &  37.42 &  1011.47 & 0.3312 & 5 & [36, 25, 25, 18, 9] \\
      1970 & & 140 & 0.2015 &  56.03 &  2284.69 & 0.3375 & 4 & [41, 38, 36, 25]    \\
      1980 & & 162 & 0.2080 &  66.96 & 12367.65 & 0.2669 & 4 & [65, 58, 20, 19]    \\
      1990 & & 169 & 0.2239 &  75.23 & 20554.50 & 0.2763 & 4 & [78, 63, 26, 2]     \\
      2000 & & 187 & 0.2833 & 105.40 & 36758.72 & 0.2585 & 4 & [88, 66, 18, 15]    \\
      \hline
\end{tabular}
  \end{center}
\caption{Characteristics of the WTW networks: $N$, number of nodes (countries); $D$, link density, i.e.\ number of edges divided by $N(N-1)$; $\langle k \rangle$, mean degree; $\langle s \rangle$, mean strength; $Q_{\max}$, maximum modularity found; $M$, number of communities; $N_{\alpha}$, sizes of the communities.}
  \label{wtw}
\end{table}

\newpage

\newpage

\end{document}